\documentclass[lettersize,journal]{IEEEtran}

\usepackage{amsmath,amsfonts}
\usepackage{algorithmic}
\usepackage{array}
\usepackage[caption=false,font=normalsize,labelfont=sf,textfont=sf]{subfig}
\usepackage{textcomp}
\usepackage{stfloats}
\usepackage{url}
\usepackage{verbatim}
\usepackage{graphicx}
\def\BibTeX{{\rm B\kern-.05em{\sc i\kern-.025em b}\kern-.08em
    T\kern-.1667em\lower.7ex\hbox{E}\kern-.125emX}}
\usepackage{balance}
\usepackage{caption}
\usepackage{siunitx}
\usepackage{tikz}
\usetikzlibrary{shapes.geometric, arrows.meta, positioning, calc, fit}

\usepackage{algorithm}
\usepackage{mathtools}	
\usepackage{amssymb}
\usepackage{amsthm}

\usepackage[utf8]{inputenc}

\usepackage{ifthen}

\usepackage{microtype}

\usepackage{caption}

\usepackage{bm}

\usepackage[draft,nomargin,marginclue,inline]{fixme}
\makeatletter
\renewcommand*\FXLayoutMarginClue[3]{%
  \marginpar[%
  \raggedleft\@fxuseface{margin}\textcolor{red}{\ignorespaces $ \Rightarrow $}]{%
    \raggedright\@fxuseface{margin}\textcolor{red}{\ignorespaces $ \Leftarrow $}}}
\makeatother	

\usepackage{tumcolor}

\usepackage{pgfplotstable}	

\pgfplotsset{
	discard if/.style 2 args={
        x filter/.append code={
            \edef\tempa{\thisrow{#1}}
            \edef\tempb{#2}
            \ifx\tempa\tempb
                
            \fi
        }
    },
    discard if not/.style 2 args={
        x filter/.append code={
            \edef\tempa{\thisrow{#1}}
            \edef\tempb{#2}
            \ifx\tempa\tempb
            \else
                
            \fi
        }
    }
}

\usepackage{cite}

\usepackage[acronym,shortcuts]{glossaries}
\newacronym{cnn}{CNN}{convolutional neural network}
\newacronym{ula}{ULA}{uniform linear array}

\usepackage{cleveref}

\usepackage{enumitem}

\tikzset{algorithm1/.style={mark options={solid},color=TUMBeamerBlue, line width=\lineWidth, mark=square, dashed}}

\pgfplotscreateplotcyclelist{miko}{%
{color=TUMBeamerRed, dash pattern=on 5pt off 1pt on 1pt off 1pt, line width=1.5pt},%
{color=TUMBeamerOrange, dash pattern=on 5pt off 2pt, line width=1.5pt,
mark=o},%
{color=black, dash pattern=on 5pt off 2pt on 3pt off 2pt, line width=1.5pt},%
{color=TUMDarkerBlue, line width=1.5pt},%
{color=black, dotted, line width=1.5pt},%
{color=TUMBeamerGreen, dotted, line width=1.5pt},%
{color=TUMBeamerYellow, dotted, line width=1.5pt},%
{color=TUMBeamerDarkRed, dotted, line width=1.5pt},%
{color=TUMBeamerLightGreen, dotted, line width=1.5pt},%
{color=TUMIvory, dotted, line width=1.5pt},%
{color=TUMLightBlue, dotted, line width=1.5pt},%
}

\DeclareMathOperator*{\argmax}{arg\,max}
\DeclareMathOperator*{\argmin}{arg\,min}

\DeclareMathOperator{\expec}{E}

\DeclareMathOperator{\tr}{tr}

\DeclareMathOperator{\uvect}{unvec}
\DeclareMathOperator{\vect}{vec}

\newcommand*{\mc}[1]{\mathcal{#1}}	

\newcommand{\calN}{\mathcal{N}}
\newcommand{\calO}{\mathcal{O}}

\newcommand*{\C}{\mathbb{C}}

\newcommand{\herm}{{\operatorname{H}}}
\newcommand{\tp}{{\operatorname{T}}}

\definecolor{myblue}{RGB}{30, 100, 200}

\newlength{\leftstackrelawd}
\newlength{\leftstackrelbwd}
\def\leftstackrel#1#2{\settowidth{\leftstackrelawd}%
	{${{}^{#1}}$}\settowidth{\leftstackrelbwd}{$#2$}%
	\addtolength{\leftstackrelawd}{-\leftstackrelbwd}%
	\leavevmode\ifthenelse{\lengthtest{\leftstackrelawd>0pt}}%
	{\kern-.5\leftstackrelawd}{}\mathrel{\mathop{#2}\limits^{#1}}}

\newcommand{\mbA}{\bm{A}}
\newcommand{\mbB}{\bm{B}}
\newcommand{\mbC}{\bm{C}}
\newcommand{\mbD}{\bm{D}}

\newcommand{\mbG}{\bm{G}}
\newcommand{\mbH}{\bm{H}}

\newcommand{\mbM}{\bm{M}}
\newcommand{\mbN}{\bm{N}}

\newcommand{\mbP}{\bm{P}}
\newcommand{\mbQ}{\bm{Q}}

\newcommand{\mbS}{\bm{S}}
\newcommand{\mbT}{\bm{T}}
\newcommand{\mbU}{\bm{U}}
\newcommand{\mbV}{\bm{V}}
\newcommand{\mbW}{\bm{W}}
\newcommand{\mbX}{\bm{X}}
\newcommand{\mbY}{\bm{Y}}
\newcommand{\mbZ}{\bm{Z}}
\newcommand{\mba}{\bm{a}}

\newcommand{\mbh}{\bm{h}}

\newcommand{\mbn}{\bm{n}}

\newcommand{\mbp}{\bm{p}}

\newcommand{\mbs}{\bm{s}}
\newcommand{\mbt}{\bm{t}}

\newcommand{\mbx}{\bm{x}}
\newcommand{\mby}{\bm{y}}

\newcommand{\mbSigma}{{\bm{\Sigma}}}

\newcommand{\mbmu}{{\bm{\mu}}}

\newcommand{\mbzero}{{\bm{0}}}

\newcommand{\hhat}{\hat{\mbh}}

\newcommand{\covhi}{\mbC_i}
\newcommand{\covhk}{\mbC_k}
\newcommand{\meanhi}{\mbmu_i}
\newcommand{\meanhk}{\mbmu_k}

\usetikzlibrary{intersections,pgfplots.fillbetween,patterns,spy}

\Crefname{figure}{Fig.}{Figs.}
\pgfplotsset{compat=1.15}

\newacronym{AWGN}{AWGN}{additive white Gaussian noise}
\newacronym{BLMMSE}{BLMMSE}{Bussgang LMMSE}
\newacronym{BS}{BS}{base station}
\newacronym{CDF}{CDF}{cumulative distribution function}
\newacronym{CNN}{CNN}{convolutional neural network}
\newacronym{CSI}{CSI}{channel state information}
\newacronym{CSIT}{CSIT}{channel state information at the transmitter}
\newacronym{DFT}{DFT}{discrete Fourier transform}
\newacronym{DL}{DL}{downlink}
\newacronym{DNN}{DNN}{deep neural network}
\newacronym{DoA}{DoA}{direction of arrival}
\newacronym{EM}{EM}{expectation maximization}
\newacronym{FDD}{FDD}{frequency division duplex}
\newacronym{GMM}{GMM}{Gaussian mixture model}
\newacronym{LMMSE}{LMMSE}{linear minimum mean square error}
\newacronym{LOS}{LOS}{line of sight}
\newacronym{LS}{LS}{least squares}
\newacronym{MSE}{MSE}{mean squared error}
\newacronym{MIMO}{MIMO}{multiple-input multiple-output}
\newacronym{MPC}{MPC}{multi-path component}
\newacronym{MT}{MT}{mobile terminal}
\newacronym{NLOS}{NLOS}{non-line of sight}
\newacronym{NN}{NN}{neural network}
\newacronym{O2I}{O2I}{outdoor-to-indoor}
\newacronym{OMP}{OMP}{orthogonal matching pursuit}
\newacronym{PDF}{PDF}{probability density function}
\newacronym{PGA}{PGA}{projected gradient ascent}
\newacronym{PSD}{PSD}{power spectral density}
\newacronym{SNR}{SNR}{signal-to-noise ratio}
\newacronym{TDD}{TDD}{time division duplex}
\newacronym{UL}{UL}{uplink}
\newacronym{ULA}{ULA}{uniform linear array}
\newacronym{URA}{URA}{uniform rectangular array}
\newacronym{UMa}{UMa}{urban macrocell}
\newacronym{nSE}{nSE}{normalized spectral efficiency}
\newacronym{cCDF}{cCDF}{complementary cumulative distribution function}
\newacronym{MU-MIMO}{MU-MIMO}{multi-user MIMO}
\newacronym{MU-MISO}{MU-MISO}{multi-user MISO}
\newacronym{BD}{BD}{block diagonalization}
\newacronym{RBD}{RBD}{regularized block diagonalization}
\newacronym{RCI}{RCI}{regularized channel inversion}
\newacronym{WMMSE}{WMMSE}{weighted minimum mean square error}
\newacronym{SWMMSE}{SWMMSE}{stochastic WMMSE}
\newacronym{SVD}{SVD}{singular value decomposition}
\newacronym{SR}{SR}{sum-rate}
\newacronym{CME}{CME}{conditional mean estimator}
\newacronym{ML}{ML}{machine learning}
\newacronym{FLOPS}{FLOPS}{floating-point operations}

\newcommand{\Ncbentries}{K}
\newcommand{\Nrx}{N_{\mathrm{rx}}}
\newcommand{\Ntx}{N_{\mathrm{tx}}}
\newcommand{\Ntxv}{N_{\mathrm{tx,v}}}
\newcommand{\Ntxh}{N_{\mathrm{tx,h}}}
\newcommand{\Krx}{K_{\mathrm{rx}}}
\newcommand{\Ktx}{K_{\mathrm{tx}}}

\begin{document}
\title{A Versatile Low-Complexity Feedback Scheme \\ for FDD Systems via Generative Modeling}
\author{ Nurettin~Turan,~\IEEEmembership{Graduate Student Member,~IEEE,} Benedikt~Fesl,~\IEEEmembership{Graduate Student Member,~IEEE,} Michael~Koller, Michael Joham,~\IEEEmembership{Member,~IEEE} and Wolfgang~Utschick,~\IEEEmembership{Fellow,~IEEE}

\thanks{Preliminary results have been presented at Asilomar'22~\cite{TuKoFeBaXuUt22}.}
\thanks{This work was partly funded by Huawei Technologies Düsseldorf GmbH, Munich, Germany.}
\thanks{The authors are with the TUM School of Computation, Information and Technology, Technische Universität München, Munich, Germany, e-mail: \{nurettin.turan, benedikt.fesl, michael.koller, joham, utschick\}@tum.de}
\thanks{This article has been accepted for publication in IEEE Transactions on Wireless Communications. This is the author's version which has not been fully edited and content may change prior to final publication. Citation information: DOI 10.1109/TWC.2023.3330902 \\
\copyright IEEE. Personal use is permitted, but republication/redistribution requires IEEE permission.
See https://www.ieee.org/publications/rights/index.html for more information.
}
}

\maketitle

\begin{abstract}
We propose a versatile feedback scheme for both single- and multi-user \ac{MIMO} \ac{FDD} systems.
Particularly, we propose utilizing a \ac{GMM} with a reduced number of parameters for codebook construction, feedback encoding, and precoder design.
The \ac{GMM} is fitted offline at the \ac{BS} to uplink training samples to approximate the channel distribution of all possible \acp{MT} within the \ac{BS} cell.
Subsequently, a codebook is constructed, with each element based on one \ac{GMM} component.
Extracting directional information from the codebook or exploiting the \ac{GMM}'s sample generation ability facilitates joint precoder design for a multi-user \ac{MIMO} system using state-of-the-art precoding algorithms.
After offloading the \ac{GMM} to the \acp{MT}, they can easily determine their feedback by selecting the index of the \ac{GMM} component with the highest responsibility for their received pilot signal.
This strategy exhibits low complexity and supports parallelization.
Simulations demonstrate that the proposed approach outperforms conventional methods, which either estimate the channel and utilize a Lloyd codebook or use a deep neural network to determine the feedback in terms of spectral efficiency or sum-rate.
The performance gains can be exploited to deploy systems with fewer pilots or feedback~bits.
\end{abstract}


\glsresetall

\section{Introduction}

In \ac{MIMO} communication systems, \ac{CSI} has to be acquired at the \ac{BS} in regular time intervals.
In \ac{FDD} mode, the \ac{BS} and the \ac{MT} transmit in the same time slot but at different frequencies. 
This breaks the reciprocity between the instantaneous \ac{UL} \ac{CSI} and \ac{DL} \ac{CSI}.
Accordingly, acquiring \ac{DL} \ac{CSI} in \ac{FDD} operation is difficult \cite{2019massive}.
The most common solution is to avoid direct feedback of the \ac{CSI} by using only a small number of feedback bits, i.e., limited feedback systems are considered \cite{Love, LaYoCh04, RaJi08}.

In conventional approaches, firstly the \ac{DL} \ac{CSI} is estimated at the \ac{MT} and subsequently the feedback information is determined.
For instance, the feedback can be used as an index for a predefined codebook of precoders or can represent quantized information (channel directions) about the \ac{DL} \ac{CSI} \cite{Love, LaYoCh04, RaJi08}.
Thus, conventional methods heavily rely on accurate \ac{CSI} estimation. 
To obtain accurate \ac{DL} \ac{CSI} at the \acp{MT}, many pilots typically need to be sent from the \ac{BS} to the \acp{MT}.
However, in massive \ac{MIMO} \ac{FDD} systems with typically many antenna elements at the \ac{BS}, the pilot overhead to fully illuminate the channel is unaffordable \cite{BjLaMa16}.
Therefore, algorithms that yield relatively good system performances with a low pilot overhead, i.e., in cases where the number of pilots is less than the number of transmit antennas, are of great interest.
It is desired to potentially circumvent explicit \ac{DL} \ac{CSI} estimation at the \ac{MT} but instead directly infer the feedback information from pilot observations.

In this context, in recent work \cite{JaLeHwReLe20, TuKoBaXuUt21, SoAtYu21, LiLiChLeShLu21, GuWeChJi22, KoSoMi21, JaLeKiLe22}, a variety of end-to-end \ac{DNN} techniques have been proposed, which process the pilot observations through neural network modules to a feedback information.
In particular, in \cite{JaLeHwReLe20}, a \ac{DNN} is employed in order to determine the feedback information for a single stream transmission in a single-user \ac{MIMO} system, that outperforms conventional approaches. 
The work in \cite{TuKoBaXuUt21} also uses a \ac{DNN} for feedback encoding but supports transmissions over multiple streams with a single \ac{DNN} for all \ac{SNR} values.
A similar approach was used in \cite{SoAtYu21} for the multi-user case with single-antenna \acp{MT}.
In \cite{LiLiChLeShLu21}, a variational autoencoder was used to provide the \ac{BS} statistical information of each \ac{MT} in combination with a stochastic iterative precoding algorithm to jointly design the precoders, again for the multi-user case with single-antenna \acp{MT}.
More recently, it was investigated in \cite{GuWeChJi22} how existing feedback schemes can be leveraged in an end-to-end limited feedback approach for the multi-user single-antenna case.
An extension to the \ac{MU-MIMO} case, i.e., \acp{MT} with multiple antennas, was proposed in 
\cite{KoSoMi21}.
Therein, the authors considered the joint design of the feedback and the precoders but compared their approach exclusively to non-iterative precoding techniques.
Another end-to-end \ac{DNN}-based approach was proposed in \cite{JaLeKiLe22}, where sub-modules of the \ac{DNN} were inspired by state-of-the-art iterative precoding algorithms.

Although the aforementioned end-to-end \ac{DNN} approaches in \cite{JaLeHwReLe20, TuKoBaXuUt21, SoAtYu21, LiLiChLeShLu21, GuWeChJi22, KoSoMi21, JaLeKiLe22} are optimized for a particular setting, they face some challenges which may potentially hinder their application in practical systems.
Specifically, the \ac{DNN} approaches are inflexible and allow support only for either the single- or multi-user mode, i.e., different task-dependent \acp{DNN} are required for each mode.
This includes the adaptation to different \ac{SNR} values, number of pilots, and number of users which usually needs additional and specifically trained networks.
Moreover, the number of \ac{DNN} parameters in \cite{JaLeHwReLe20, TuKoBaXuUt21, SoAtYu21, LiLiChLeShLu21, GuWeChJi22, KoSoMi21, JaLeKiLe22} scales quadratically with the product of transmit and receive antennas, leading to difficulties in the training time and the convergence abilities. 
In fact, in \cite{KoSoMi21, JaLeKiLe22}, only relatively small antenna configurations are considered which are not in accordance with trends towards massive \ac{MIMO}.
Even more disadvantageously, the offload amount in order to communicate the \ac{DNN} parameters from the \ac{BS} to the \acp{MT} is drastically increasing in the number of antennas, the supported \ac{SNR} range, transmission modes, and for a varying number of pilots, resulting in unaffordable signaling overhead. 

\Ac{ML} techniques typically require a representative dataset of channels stemming from the \ac{BS} cell for their training phase.
In \ac{FDD} mode, the \acp{MT} would have to collect large amounts of \ac{DL} \ac{CSI} and either need to perform the training themselves or to share the collected data with the \ac{BS}.
The corresponding computation and signaling overhead is generally unaffordable in practice. 
Recently, in \cite{utschick2021}, it has been shown that \ac{DL} \ac{CSI} training data can be replaced with \ac{UL} \ac{CSI} training data even for the design of \ac{DL} functionalities.
This completely eliminates the aforementioned overhead.
The \ac{UL} \ac{CSI} can be acquired at the \ac{BS} during the regular \ac{UL} transmission.
The observation in \cite{utschick2021} was confirmed for various DL functionalities, e.g., in~\cite{fesl2021centralized, BaRiAlUt23, TuKoBaXuUt21, TuKoRiFeBaXuUt21}.
Consequently, in this work, we also utilize the idea of centrally learning \ac{DL}-related functionalities at the \ac{BS} using \ac{UL} training data.

\acp{GMM} are widely adopted in the wireless communications literature.
For example, in \cite{MuMiDe20}, \cite{LiZhMaZh20}, and \cite{GuZhYi19}, \acp{GMM} are used for predicting channel states, multi-path clustering, and pilot optimization, respectively.
In \cite{KoFeTuUt21J}, a \ac{GMM} is used to approximate the true but unknown channel \ac{PDF} and a powerful channel estimator is derived.
The strong performance is justified by the universal approximation ability of \acp{GMM}, cf. \cite{NgNgChMc20}.
The primary motivations for leveraging \acp{GMM} in this work to propose a versatile low-complexity feedback scheme for point-to-point \ac{MIMO} and \ac{MU-MIMO} \ac{FDD} systems, apart from their universal approximation property, are the following. 
On the one hand, \acp{GMM} comprise a discrete latent space, which enables the clustering of channels and makes the inference of the latent variable, given an observation, tractable.
These properties are exploited to design a novel codebook and propose a limited feedback scheme. 
On the other hand, \acp{GMM} are generative models. 
Generative models refer to techniques that aim to learn the underlying distribution of a training data set with the goal to enable the generation of new samples that resemble the original distribution.
We propose to utilize this sample generation ability in combination with the feedback information to enhance the precoder design via a stochastic algorithm.
In recent years, other generative concepts such as generative adversarial networks \cite{GoPoMiXuWaOzCoBe14} and variational autoencoders \cite{DiMa14} also gained a lot of attention.
In the context of wireless communications, these generative models were utilized for, e.g., channel estimation~\cite{BaDoJaDiAn21}, precoding~\cite{LiLiChLeShLu21}, and as a channel modeling framework~\cite{YaLiZhQiZhWa19, YeLiLiJu20}.

The contributions of this work are summarized as follows:
\begin{enumerate}
    \item The \ac{GMM} can be centrally fitted at the \ac{BS} using solely \ac{UL} training data. We propose to cluster the training data according to the \ac{GMM} components and design a codebook entry per \ac{GMM} component, which yields a scenario-specific codebook and supports the single-user mode.
    By offloading the \ac{GMM} to the \acp{MT} upon entering the coverage area of the \ac{BS}, we propose to use the index, which represents the \ac{GMM} component that yields the highest responsibility of the observed pilot signal of a \ac{MT} as feedback information. Thereby, the responsibility evaluates the probability that the channel to a particular \ac{MT} stems from the corresponding \ac{GMM} component.
    \item We further propose two approaches to support the multi-user mode.
    By extracting directional information as relevant features of the constructed codebook, jointly designing precoders for a \ac{MU-MIMO} system with an arbitrary precoding algorithm, cf. \cite{SpSwHa04, StHa08, PeHoSw05, LeLeHoLe08, ChAgDeCi08, ShRaLuHe11, HuCaYuQiXuYuDi21}, is possible. 
    Alternatively, we propose to leverage the \ac{GMM}'s sample generation ability in order to provide statistical information of each \ac{MT} to the \ac{BS} and design the precoders using a state-of-the-art stochastic iterative precoding algorithm, e.g.,~\cite{RaBoLu13, RaSaLu16}.
    Thus, the proposed scheme allows to influence the complexity of the precoder design at the \ac{BS} depending on the selected transmission mode and the precoding algorithm.
    \item The complexity of determining the feedback at the \ac{MT} side by using the \ac{GMM} does not scale with the number of transmit antennas, in contrast to conventional approaches, and even allows for parallelization.
    Due to the analytic representation of the \ac{GMM}, the feedback scheme can be straightforwardly adapted to any SNR, pilot configuration, and number of users without retraining, which is in contrast to the end-to-end \ac{DNN}-based approaches \cite{KoSoMi21, JaLeKiLe22}.
    Moreover, model-based insights can be leveraged to drastically decrease the training time and the offloading overhead, and allow to conveniently scale with larger antenna dimensions.
    Thus, the \ac{GMM}-based feedback scheme is particularly beneficial for massive \ac{MIMO} systems.
    \item Despite exhibiting a lower complexity, the proposed \ac{GMM}-based feedback scheme provides high robustness against \ac{CSI} imperfections and outperforms conventional single- and multi-user precoding approaches, especially in settings with a low pilot overhead.
    With extensive simulations, we show that the performance gains achieved with our proposed scheme can be leveraged to deploy system setups with, e.g., a reduced number of pilots or with a smaller number of feedback bits, as compared to classical approaches.
\end{enumerate}

The paper is structured as follows.
The system models are introduced in Section \ref{sec:system_channel_model}.
In Section \ref{sec:p2pmimo} and Section \ref{sec:mumimo}, we discuss conventional methods and present the proposed approaches.
In Section \ref{sec:discussion_universality}, we discuss the versatility of the proposed scheme.
In Section~\ref{sec:baseline_channel_estimators}, channel estimators are discussed and in Section \ref{sec:comp_ana} a complexity analysis is provided.
Simulation results are provided in Section \ref{sec:sim_results}, and in Section \ref{sec:conclusion} conclusions are drawn.

\emph{Notation:}
Matrices and vectors are denoted with bold uppercase and bold lowercase letters, respectively.
The transpose or conjugate transpose of a matrix $\mbA$ is denoted by $\mbA^\tp$ or $\mbA^\herm$, respectively.
The all-zeros vector and the identity matrix with appropriate dimensions are denoted by $\mathbf{0}$ or $\mathbf{I}$, respectively.
The Euclidean norm of a vector \( \mba \in \C^N \) is denoted by \( \| \mba \| \).
The cardinality of a set $\mathcal{V}$ is denoted by $|\mathcal{V}|$.
A complex-valued normal distribution with mean vector \( \mbmu \) and covariance matrix \( \mbC \) is denoted by \( \calN_{\C}(\mbmu, \mbC) \) and~$\sim$~stands for \textquotedblleft distributed as\textquotedblright.
The determinant or the trace of matrix $\mbA$ is given by $\det(\mbA)$ and $\tr(\mbA)$, respectively.
The vectorization (stacking columns) of a matrix \( \mbA \in \C^{m\times N} \) is written as \( \mba=\vect(\mbA) \in \C^{mN} \), and the reverse operation is denoted by \(\mbA = \uvect(\mba)\).
The Kronecker product of two matrices \( \mbA \in \C^{m_1\times N_1} \) and \( \mbB \in \C^{m_2\times N_2} \) is \( \mbA \otimes \mbB \in \C^{m_1m_2\times N_1N_2} \).

\section{System and Channel Models}
\label{sec:system_channel_model}

\subsection{Data Transmission -- Point-to-Point MIMO System}

The \ac{DL} received signal of a point-to-point \ac{MIMO} system can be expressed as $\mby^\prime = \mbH \mbx + \mbn^\prime$,
where $\mby^\prime \in \C^{\Nrx}$ is the receive vector, $\mbx \in \C^{\Ntx}$ is the transmit vector sent over the \ac{MIMO} channel $\mbH \in \C^{\Nrx \times \Ntx}$, and $\mbn^\prime \sim \mathcal{N}_\C(\mathbf{0},
\sigma_n^2 \mathbf{I}_{\Nrx})$ denotes the \ac{AWGN}.
In this paper, we consider configurations with $\Nrx < \Ntx$. The \ac{BS} is equipped with a \ac{URA} and the \ac{MT} is equipped with a \ac{ULA}.
If perfect \ac{CSI} is known to both, the transmitter and receiver, and assuming transmit data with zero-mean Gaussian distribution, the capacity of the \ac{MIMO} channel is $ C = \max_{\mbQ \succeq \mbzero, \tr\mbQ\leq\rho} \log_2 \det\left( \mathbf{I} + \frac{1}{\sigma_n^2} \mbH \mbQ \mbH^\herm\right)$, e.g.,~\cite[page~326]{Goldsmith}, where $\mbQ \in \C^{\Ntx \times \Ntx}$ is the transmit covariance matrix, $\rho$ is the transmit power, and the transmit vector is given by \( \mbx = \mbQ^{1/2} \mbs \) with \( \expec[\mbs \mbs^\herm] = \mathbf{I}_{\Ntx} \)~\cite{Love}.
The optimal transmit covariance matrix $\mbQ^\star$ achieves the capacity and can be obtained by decomposing the channel into $\Nrx$ parallel streams and employing water-filling \cite{Telatar99capacityof}. 
Since channel reciprocity does not hold in \ac{FDD} systems, only the \ac{MT} could compute the optimal transmit covariance matrix \( \mbQ^\star \) if the \ac{DL} \ac{CSI} is estimated perfectly. 
This makes some form of feedback from the \ac{MT} to the \ac{BS} necessary.
Ideally, the \ac{MT} would feed the complete \ac{DL} \ac{CSI} back to the \ac{BS}, which is infeasible in general.
Instead, a small number of \( B \) bits is fed back to the \ac{BS}.
The \( B \) feedback bits are commonly used for encoding an index $k^\star \in \{1,2, \cdots, 2^B\}$ that specifies an element from a set of $ 2^B $ pre-computed transmit covariance matrices $ \mathcal{Q} = \{\mbQ_1, \mbQ_2, \dots, \mbQ_{2^B} \} $.
Finally, the \ac{BS} employs the transmit covariance matrix \( \mbQ_{k^\star} \) for data transmission \cite{Love}.

\subsection{Data Transmission -- Multi-user MIMO System}

We consider a single-cell \ac{MU-MIMO} system in the \ac{DL}, where linear precoding is adopted.
The system consists of a \ac{BS} equipped with $\Ntx$ transmit antennas and $J$ \acp{MT}.
Each \ac{MT} $j \in \mathcal{J} = \{1, 2, \dots, J\}$ is equipped with $\Nrx$ antennas.
Let the transmit signal vector corresponding to \ac{MT} $j$ be $\mbs_j \in \C^{d_j}$, where $d_j$ is the number of data streams.
We assume that $\expec[\mbs_j]=\mathbf{0}$ and $\expec[\mbs_j\mbs_j^\herm]=\mathbf{I}_{d_j}$. 
Furthermore, the symbols sent to each \ac{MT} are assumed to be independent of each other.
The overall precoded \ac{DL} data vector is $\mbx = \sum_{j=1}^{J}\mbM_j\mbs_j$ where $\mbM_j \in \C^{\Ntx \times d_j}$ is the precoding matrix applied at the \ac{BS} to process the transmit signal of \ac{MT} $j$ (without loss of generality $d_j=\Nrx,  \forall j$ is set in the following, if not mentioned otherwise).
The precoders satisfy the transmit power constraint $\operatorname{tr}(\sum_{j=1}^J\mbM_j \mbM_j^\herm) = \rho$.
Thus, the received signal at \ac{MT} $j$ is
\begin{equation} 
    \mby^\prime_j = \mbH_j \mbM_j \mbs_j + \sum_{m=1, m\neq j}^{J} \mbH_j\mbM_m\mbs_m+\mbn^\prime_j, \ \forall j \in \mathcal{J}
\end{equation}
where $\mbH_j \in \C^{\Nrx \times \Ntx}$ is the MIMO channel from the \ac{BS} to \ac{MT} $j$ and $\mbn^\prime_j \in \C^{\Nrx} \sim \mathcal{N}_\C(\bm{0}, \sigma_j^2 \mathbf{I}_{\Nrx})$ denotes the AWGN of \ac{MT} $j$.
The instantaneous achievable rate of \ac{MT} $j$ can be written as 
\begin{multline} 
\label{eq:inst_sumrate}
    R_j^{\text{inst}} = \log_2 \det \Big(\mathbf{I} + \mbH_j\mbM_j\mbM_j^\herm \mbH_j^\herm \\
    \times \big(\sum_{m\neq j} \mbH_j \mbM_m \mbM_m^\herm \mbH_j^\herm + \sigma_j^2 \mathbf{I} \big)^{-1} \Big).  
\end{multline}

If the \ac{BS} had access to the perfect \ac{DL} \ac{CSI} of each of the \acp{MT}, it could employ common non-iterative algorithms such as \ac{BD} \cite{SpSwHa04}, \ac{RBD} \cite{StHa08}, or \ac{RCI} \cite{PeHoSw05, LeLeHoLe08}, or iterative algorithms such as the iterative \ac{WMMSE} algorithm \cite{ChAgDeCi08, ShRaLuHe11, HuCaYuQiXuYuDi21}, in order to jointly design the precoders $\mbM_j$ of all \acp{MT}, $j \in \mathcal{J}$. 
However, for the considered limited feedback, each \ac{MT} $j$ is assumed to encode an index $k_j^\star$ with \( B \) bits, representing quantized information regarding the \ac{DL} \ac{CSI}, and feeds this information back to the \ac{BS}.

In the seminal work \cite{RaJi08}, such a limited feedback system was investigated, where quantized information regarding the \ac{CSI} of each \ac{MT} is fed back to the \ac{BS} after determining the best entry of a randomly generated \ac{MT}-specific codebook.
The random quantization codebook of each \ac{MT} is assumed to be perfectly known to the \ac{BS} \cite{RaJi08}.
Then, \ac{BD} was employed at the \ac{BS} in order to jointly design the precoders.
Multi-user systems with $J=\frac{\Ntx}{\Nrx}\geq2$ were considered, i.e., the total number of receive antennas $J\Nrx$ equals the number of transmit antennas $\Ntx$, in order to omit having to select a subset of \acp{MT} for transmission~\cite{RaJi08}. 
We will also consider such setups in our simulations.

\subsection{Pilot Transmission Phase}

In the pilot transmission phase, the \ac{DL} received signal of each \ac{MT} $j \in \mathcal{J}$ is
\begin{equation} \label{eq:noisy_obs}
    \mbY_j = \mbH_j \mbP + \mbN_j \in \C^{\Nrx \times n_p}
\end{equation}
where $n_p$ is the number of transmitted pilots and $\mbN_j = [\mbn^{\prime}_{j,1}, \mbn^{\prime}_{j,2}, \dots, \mbn^{\prime}_{j,n_p}] \in \C^{\Nrx \times n_p}$ with $\mbn^\prime_{j,l} \sim \mathcal{N}_\C(\bm{0}, \sigma_j^2 \mathbf{I}_{\Nrx}), \forall l$. The pilot matrix $\mbP \in \C^{\Ntx \times n_p}$ is a $2$D-DFT (sub)matrix, constructed by the Kronecker product of two \ac{DFT} matrices, $\mbP = \mbP_\text{h} \otimes \mbP_\text{v}$, where each column $\mbp_p$ of $\mbP$, for $p \in \{1,2, \dots, n_p\} $, is normalized such that $\|\mbp_p\|^2=\rho$ to fulfill the power constraint, since we employ a \ac{URA} at the \ac{BS}, see e.g., \cite{TsZhWa18}. In this work, we consider $n_p\leq \Ntx$, i.e., the number of pilots is less than or equal to the number of transmit antennas.
For what follows, it is convenient to vectorize~\eqref{eq:noisy_obs}, yielding $\mby_j = \mbA \mbh_j + \mbn_j$, with the definitions \( \mbh_j = \vect(\mbH_j) \), \( \mby_j = \vect(\mbY_j) \), \( \mbn_j = \vect(\mbN_j) \), \( \mbA = \mbP^\tp \otimes \mathbf{I}_{\Nrx} \) and $\mbn_j \sim \mathcal{N}_\C(\mathbf{0}, \mbSigma)$ with $\mbSigma = \sigma_n^2 \mathbf{I}_{\Nrx n_p}$.
In case of a point-to-point \ac{MIMO} system, we drop the index $j$ for notational convenience and end up with
\begin{equation} \label{eq:noisy_obs_p2p}
    \mby = \mbA \mbh + \mbn \in \C^{\Nrx n_p}.
\end{equation}

\subsection{Channel Model and Data Generation} \label{sec:data_generation}

The QuaDRiGa channel simulator \cite{QuaDRiGa1, QuaDRiGa2} is used to generate \ac{CSI} for the \ac{UL} and \ac{DL} domains in an \ac{UMa} scenario.
The carrier frequencies are $\SI{2.53}{\giga\hertz}$ for the \ac{UL} and $\SI{2.73}{\giga\hertz}$ for the \ac{DL}, such that there is a frequency gap of $\SI{200}{\mega\hertz}$.
The \ac{BS} uses a \ac{URA} with ``3GPP-3D'' antennas and the \acp{MT} use \acp{ULA} with ``omni-directional'' antennas.
The \ac{BS} covers a $\SI{120}{\degree}$ sector and is placed at $\SI{25}{\meter}$ height.
The minimum and maximum distances between \acp{MT} and the \ac{BS} are $\SI{35}{\meter}$ and $\SI{500}{\meter}$, respectively.
In $80 \ \%$ of the cases, the \acp{MT} are located indoors at different floor levels.
The outdoor \acp{MT} have a height of $\SI{1.5}{\meter}$.
A QuaDRiGa \ac{MIMO} channel is given by $\mbH = \sum_{\ell=1}^{L} \mbG_{\ell} e^{-2\pi j f_c \tau_{\ell}}$
with $\ell$ being the path number, $L$ the number of \acp{MPC}, $f_c$ the carrier frequency, and $\tau_{\ell}$ the \( \ell \)th path delay.
The number \( L \) depends on whether there is \ac{LOS}, \ac{NLOS}, or \ac{O2I} propagation: $L_\text{LOS} = 37$, $L_\text{NLOS} = 61$, or $L_\text{O2I} = 37$~\cite{QuaDRiGa2}.
The coefficients matrix $\mbG_{\ell}$ consists of one complex entry for each antenna pair and comprises the attenuation of a path, the antenna radiation pattern weighting, and the polarization.
The generated channels are post-processed to remove the path gain \cite{QuaDRiGa2}.
In the following, we denote by
\begin{equation} \label{eq:H_dataset}
     \mathcal{H} = \{\mbh_m = \vect{(\mbH_m)}\}_{m=1}^{M}  
\end{equation}
the training dataset consisting of $M$ channels from the scenario described above. 

\section{Point-to-Point MIMO System: \\ Codebook Design \& Feedback Encoding}
\label{sec:p2pmimo}

\subsection{Conventional Codebook Construction and Encoding Scheme} \label{sec:conventionalscheme}

In an offline training phase, one can construct a codebook $\mathcal{Q}$ with \( K = 2^B \) elements.
A standard codebook construction approach uses Lloyd's algorithm~\cite{LiBuGr80, LaYoCh04}.
Given a training dataset of channels \( \mathcal{H} \), see \eqref{eq:H_dataset}), the iterative Lloyd clustering algorithm alternates between two stages until a convergence criterion is met. 
Note that we use the channel matrix $\mbH$ and its vectorized expression $\mbh$ interchangeably in the following for ease of notation.
We write $ \{ \mbQ_k^{(i)} \}_{k=1}^{\Ncbentries} $ for the codebook in iteration $ i $.
The two stages in iteration $ i $ are:
\begin{enumerate}
    \item Divide the training dataset $ \mathcal{H} $ into $ \Ncbentries $ clusters $ \mathcal{V}_k^{(i)} $:
    \begin{equation}\label{eq:lloyd_stage_1_conv}
        \mathcal{V}_k^{(i)} = \{ \mbh \in \mc{H} \mid r(\mbH, \mbQ_k^{(i)}) \geq r(\mbH, \mbQ_j^{(i)}) \text{ for } j\neq k \}.
    \end{equation}
    \item Update the codebook:
    \begin{align}\label{eq:lloyd_stage_2}
        &\mbQ_k^{(i+1)} = \argmax_{\mbQ \succeq \mbzero} \frac{1}{|\mathcal{V}_k^{(i)}|} \sum_{\vect(\mbH)\in\mathcal{V}_k^{(i)}} r(\mbH,\mbQ) \\
        & \text{subject to} \quad \operatorname{tr}(\mbQ) \leq \rho \nonumber
    \end{align}
\end{enumerate}
where for a channel matrix \( \mbH \) and a covariance matrix \( \mbQ \),
\begin{equation}
    r(\mbH, \mbQ) = \log_2 \det\left( \mathbf{I} + \frac{1}{\sigma_n^2} \mbH \mbQ \mbH^\herm\right)
    \label{speceff}
\end{equation}
is the spectral efficiency.
The optimization problem in stage 2) is solved via \ac{PGA}, cf.~\cite{TuKoBaXuUt21, HuScJoUt08}.
To initialize the algorithm, stage 1) is replaced with a random partition of \( \mc{H} \) in the first iteration.

\emph{Lau's heuristic} \cite{LaYoCh04}: In order to avoid solving the costly optimization problem in stage 2) in every iteration, a heuristic for the codebook update is given in \cite{LaYoCh04}: A representative matrix
$\mbS_k^{(i)} = \frac{1}{|\mathcal{V}_k^{(i)}|} \sum_{\vect(\mbH)\in\mathcal{V}_k^{(i)}} \mbH^\herm \mbH$ is calculated for every cluster \( \mc{V}_k^{(i)} \), and then the matrices $\mbS_k^{(i)}$ are decomposed into $\Nrx$ parallel streams and water-filling is employed, yielding the updated codebook entries $\mbQ_k^{(i+1)}$.
However, as attested by the simulation results in \cite{TuKoFeBaXuUt22}, the heuristic approach leads to a performance loss as compared to the \ac{PGA} approach.
Thus, we will restrict our analysis to the \ac{PGA} approach in this work.

In the online phase, following the pilot transmission phase, the \ac{MT} is assumed to estimate the \ac{DL} channel \( \hat{\mbH} \) and then uses it to determine the best codebook entry \( \mbQ_{k^\star} \) of the commonly shared codebook $ \mathcal{Q}$ via:
\begin{equation}\label{eq:codebook_index_selection}
    k^\star = \argmax_{k \in \{1, \dots, 2^B\}} \log_2 \det\left( \mathbf{I} + \frac{1}{\sigma_n^2} \hat{\mbH} \mbQ_k \hat{\mbH}^\herm\right).
\end{equation}
The feedback consists of the index \( k^\star \) encoded by \( B \) bits and the \ac{BS} employs the transmit covariance matrix \( \mbQ_{k^\star} \) for data transmission.

\subsection{Proposed Codebook Construction and Encoding Scheme}

\label{sec:proposedscheme}

The channel characteristics of the whole propagation environment of a \ac{BS} cell can be described by means of a \ac{PDF} $f_{\mbh}$.
This \ac{PDF} \( f_{\mbh} \) describes the stochastic nature of all channels in the whole coverage area of the \ac{BS}.
The channel of any \ac{MT} within the \ac{BS} cell is a realization of a random variable with \ac{PDF} \( f_{\mbh} \).
The main problem is that this \ac{PDF} is typically not available analytically.
In this setting, \ac{ML} approaches play an increasingly important role.
They aim to implicitly learn the underlying \ac{PDF} from representative data samples stemming from the \ac{BS} cell, cf. \cite{GuWeChJi22, TuKoBaXuUt21}.
In contrast, motivated by universal approximation properties of \acp{GMM}~\cite{NgNgChMc20}, we fit a \ac{GMM} \( f_{\mbh}^{(K)} \) with \( K \) components in order to approximate the unknown channel \ac{PDF} \( f_{\mbh} \), similar as in \cite{KoFeTuUt21J, ToFeGrKoUt22} in an analytic form.
In this work, the training data set stems from a stochastic-geometric channel simulator \cite{QuaDRiGa1}, similarly as in \cite{GuWeChJi22, KoFeTuUt21J}.
Alternatively, training data can be acquired, for instance, from a measurement campaign \cite{ToFeGrKoUt22} or by using a ray-tracing software~\cite{BaRiAlUt23}. 
In \cite{FeTuJoUt23}, it was shown that a \ac{GMM} can be even learned from imperfect data.
The analysis with these different training data sources and data imperfections is out of the scope of this work.

A \ac{GMM} is a \ac{PDF} of the form~\cite[Subsection~2.3.9]{bookBi06}
\begin{equation}\label{eq:gmm_of_h}
    f^{(K)}_{\mbh}(\mbh) = \sum_{k=1}^K p(k) \calN_{\C}(\mbh; \mbmu_k, \mbC_k)
\end{equation}
where every summand is one of its \( K \) \textit{components}.
Maximum likelihood estimates of the parameters of a \ac{GMM}, viz., the means $\mbmu_k$, the covariances $\mbC_k$, and the mixing coefficients $p(k)$, can be computed using a training dataset \(\mathcal{H} \), see \eqref{eq:H_dataset}, and an \ac{EM} algorithm, see~\cite[Subsection~9.2.2]{bookBi06}, which can be summarized with the following four steps: \emph{i})~Initialize the parameters of the \ac{GMM} and calculate the initial value of the log-likelihood; \emph{ii})~E-step: Determine the responsibilities, which evaluate the probability that a given data point belongs to (or is explained by) a particular component of the \ac{GMM}; \emph{iii})~M-step: Update the parameters using the current responsibilities; \emph{iv})~Evaluate the log-likelihood with the updated parameters and repeat the E-step and M-step until the convergence of the log-likelihood.

After a \ac{GMM} is fitted, we can determine the likelihood that a particular channel $\mbh$ stems from one of the components by evaluating the responsibilities~\cite[Section~9.2]{bookBi06}:
\begin{equation}\label{eq:responsibilities_h}
    p(k \mid \mbh) = \frac{p(k) \calN_{\C}(\mbh; \mbmu_k, \mbC_k)}{\sum_{i=1}^K p(i) \calN_{\C}(\mbh; \mbmu_i, \mbC_i) }.
\end{equation}

Due to the joint Gaussianity of each \ac{GMM} component and the \ac{AWGN}, the approximate \ac{PDF} of the observations is straightforwardly computed using the \ac{GMM} from \eqref{eq:gmm_of_h} as 
\begin{equation}\label{eq:gmm_y}
    f_{\mby}^{(K)}(\mby) = \sum_{k=1}^K p(k) \calN_{\C}(\mby; \mbA \meanhk, \mbA \covhk \mbA^\herm + \mbSigma),
\end{equation}
which is also a \ac{GMM} (\ac{GMM} of the observation).
Since \acp{GMM} allow to calculate the responsibilities by evaluating Gaussian likelihoods, we can compute:
\begin{equation}\label{eq:responsibilities}
    p(k \mid \mby) = \frac{p(k) \calN_{\C}(\mby; \mbA \meanhk, \mbA \covhk \mbA^\herm + \mbSigma)}{\sum_{i=1}^K p(i) \calN_{\C}(\mby; \mbA \meanhi, \mbA \covhi \mbA^\herm + \mbSigma) }.
\end{equation}

The idea of the proposed method is to compute a codebook transmit covariance matrix \( \mbQ_k \) for every component of the \ac{GMM} and to use the responsibilities \( p(k\mid \mby) \) to determine the feedback index.
In detail, in an offline training phase, we take \( K = 2^B \) as the number of \ac{GMM} components, use a training dataset of channels \(\mathcal{H} \) to fit a \( K \)-components \ac{GMM} \( f_{\mbh}^{(K)} \), and compute a codebook \( \mc{Q} = \{ \mbQ_k \}_{k=1}^K \) of transmit covariance matrices---one matrix for every \ac{GMM} component---exclusively at the \ac{BS}.
We explain the codebook construction in another paragraph below.
During the online phase, when the objective is to determine a feedback index, we bypass explicit channel estimation and directly determine a feedback index using the responsibilities computed via \( \mby \):
\begin{equation} \label{eq:ecsi_index}
    k^\star = \argmax_{k } {~p(k \mid \mby)},
\end{equation}
i.e., the highest responsibility of the observed pilot signal of a \ac{MT} serves as the feedback information.
Thus, we find the feedback index \( k^\star \) without requiring (estimated) \ac{CSI}.
Note, we thereby also avoid the evaluation of the \( \log_2 \det \)  in~\eqref{eq:codebook_index_selection}.
Furthermore, the knowledge of the codebook at the \ac{MT} is not required. 
The \ac{MT} only requires the parameters of the \ac{GMM} to compute \eqref{eq:ecsi_index}.

We can think of \( p(k \mid \mby) \) as an approximation of \( p(k \mid \mbh) \) from~\eqref{eq:responsibilities_h} because of the fixed noise covariance of every component.
That is, since there is a true underlying channel \( \mbh \) leading to the current observation \( \mby = \mbA \mbh + \mbn \), the responsibility \( p(k \mid \mby) \) can be seen as an approximation of the probability \( p(k \mid \mbh) \) that the channel \( \mbh \) stems from the \( k \)th \ac{GMM} component.
To investigate the influence of using \( p(k \mid \mby) \) instead of \( p(k \mid \mbh) \), it is interesting to look at the performance of the feedback information calculated as
\begin{equation} \label{eq:pcsi_index}
    k^\star = \argmax_{k } ~{p(k \mid \mbh)}.
\end{equation}
Note that this approach is infeasible in practice because the channel \( \mbh \) would have to be known in the online phase.
Nevertheless, it serves as a baseline for the performance analysis.

\emph{Proposed codebook construction:}
Once the training dataset \( \mc{H} \) has been used to fit a \( K \)-components \ac{GMM}, we cluster the training data according to their \ac{GMM} responsibilities, i.e., channels exhibiting high similarities measured in terms of the responsibilities are assigned to the same component and thereby form a cluster.
That is, we partition \( \mc{H} \) into \( K \) disjoint sets denoted by
\begin{equation}\label{eq:lloyd_stage_1}
    \mc{V}_k = \{ \mbh \in \mathcal{H} \mid p(k \mid \mbh) \geq p(j \mid \mbh) \text{ for } k\neq j \}
\end{equation}
for \( k = 1, \dots, K \).
We now determine the codebook \( \mc{Q} = \{ \mbQ_k \}_{k=1}^K \) by computing every transmit covariance matrix \( \mbQ_k \) such that it maximizes the summed rate in \( \mc{V}_k \):
\begin{align}\label{eq:gmmcb_stage_2}
    &\mbQ_k = \argmax_{\mbQ \succeq \mbzero} \frac{1}{|\mathcal{V}_k|} \sum_{\vect(\mbH)\in\mathcal{V}_k} r(\mbH,\mbQ) \\
    & \text{subject to} \quad \operatorname{tr}(\mbQ) \leq \rho \nonumber
\end{align}
where $r(\mbH, \mbQ)$ is the spectral efficiency defined in \eqref{speceff}.
This optimization problem is solved via a \ac{PGA} algorithm similar as in Section \ref{sec:conventionalscheme}, cf.~\cite{TuKoBaXuUt21, HuScJoUt08}.
Analogously, we can replace the optimization problem in~\eqref{eq:gmmcb_stage_2} with Lau's heuristic from~\cite{LaYoCh04} (cf. \ref{sec:conventionalscheme}) to compute a transmit covariance matrix for every \ac{GMM} component, which degrades the performance \cite{TuKoFeBaXuUt22}.
Thus, we again restrict our analysis to the \ac{PGA} approach.

In summary, the \ac{GMM} is used twice: Once for codebook construction (done offline) and afterwards for the determination of a feedback index (done online). 
For the latter, it is not necessary to estimate the channel and evaluating~\eqref{eq:codebook_index_selection} is avoided.
This is particularly beneficial for the online computational complexity, which is discussed in Section \ref{sec:comp_ana}.
Moreover, the \ac{GMM} of the observations, see \eqref{eq:gmm_y}, can be straightforwardly adapted at the \ac{MT} to any \ac{SNR} and pilot configuration, by simply updating the means and covariances, cf. \eqref{eq:gmm_y}, without retraining.

\newpage
\section{Multi-User MIMO System: \\ Feedback Encoding \& Precoder Design}
\label{sec:mumimo}

\subsection{Conventional Method} \label{sec:conventional_method}

In \cite{RaJi08}, the authors considered limited feedback in the \ac{MU-MIMO} setting, where 
\ac{BD} was applied as precoding algorithm and a uniform power allocation policy was adopted, i.e., no water-filling across streams was conducted.
Accordingly, the feedback conveys information regarding the spatial direction of each \ac{MT}'s channel to the \ac{BS}, and no magnitude information is fed back.

Consider the \ac{SVD} of the (estimated) channel $\hat{\mbH}_j = \mbU_{j} \mbS_j \mbV_{j}^\herm$ of \ac{MT} $j$, where $\mbS_j$ contains the singular values in descending order on its diagonal, and let $\bar{\mbV}^\herm_{j}$ contain the first $\Nrx$ rows of $\mbV^\herm_{j}$, cf. e.g., \cite{LoHe05}. 
Given the matrix $\bar{\mbV}^\herm_{j}$ , the idea from \cite{RaJi08} is to feed back the information regarding $\bar{\mbV}^\herm_{j}$ to the \ac{BS} by using a random quantization codebook \cite{SaHo04}.
In particular, each \ac{MT}-specific random quantization codebook is fixed beforehand and is known to the \ac{BS}. 
That is, the codebook $\mc{Q} = \{\mbW_1, \mbW_2, \cdots, \mbW_{K}\}$ of a particular \ac{MT} consists of $K=2^B$ sub-unitary matrices (i.e., $\mbW_k^\herm \mbW_k = \mathbf{I}_{\Nrx}$, for \( k = 1, \dots, K \)) which are chosen independently and are uniformly distributed over the Grassmann manifold \cite{RaJi08, DaLiRi08}.
The elements of the random quantization codebook are thus constructed by generating an $\Ntx \times \Nrx$ dimensional matrix with i.i.d. complex Gaussian entries and then computing an $\Ntx \times \Nrx$ dimensional subspace spanned by the matrix using the procedure described in \cite{Me07}.
The selection method considered in \cite{RaJi08} was the chordal distance metric, i.e., $k^\star_j =  \argmin_{k} \left(  {\Nrx - \operatorname{tr}(\bar{\mbV}_{j}^\herm \mbW_k \mbW_k^\herm \bar{\mbV}_{j})} \right)$, and other metrics were not investigated in \cite{RaJi08}.
We also considered the chordal distance in our simulations but found that using a capacity inspired selection metric (like in \cite{LoHe05}) performed consistently better. Thus, we used the same evaluation principle per \ac{MT} as in \eqref{eq:codebook_index_selection}, but replaced the transmit covariance matrices by $\frac{\rho}{\Nrx}\mbW_k \mbW_k^\herm$:
\begin{equation}\label{eq:codebook_index_selection_j}
    k^\star_j = \argmax_{k }~\log_2 \det\left( \mathbf{I} + \frac{\rho}{\sigma_n^2 \Nrx} \hat{\mbH}_j \mbW_k \mbW_k^\herm \hat{\mbH}_j^\herm\right).
\end{equation}
In this way, we also account for the \ac{SNR} compared to the chordal distance metric, which does not depend on the \ac{SNR}.
For the sake of brevity, we will not show the results obtained by using the chordal distance metric in Section \ref{sec:sim_results}, since we compare to the consistently better baseline.

Each \ac{MT} reports $k^\star_j$ to the \ac{BS} and the \ac{BS} then represents each \ac{MT}'s channel by $\widetilde{\mbH}_j = \mbW^\herm_{k^\star_j}$. 
Given the quantized directional information regarding each \ac{MT}'s channel, the \ac{BS} can employ
a common precoding algorithm in order to jointly design the precoders.
The authors of \cite{RaJi08} focused their analysis on \ac{BD} and adopted the uniform power allocation policy.
In contrast, we will consider a broad range of precoding algorithms, including non-iterative algorithms, \ac{RBD} with the uniform power allocation policy \cite{StHa08} (an extension of \ac{BD}), \ac{RCI} \cite{LeLeHoLe08}, and the iterative \ac{WMMSE} algorithm \cite[Algorithm~1]{HuCaYuQiXuYuDi21}, in order to jointly design the precoders $\mbM_j$ of all \acp{MT}~$j \in \mathcal{J}$.

\subsection{Proposed Subspace-based Method} \label{sec:sub_method}

Inspired by the approach from \cite{RaJi08}, we propose to use the following approach to obtain quantized information regarding each \ac{MT}'s channel.
Each transmit covariance matrix of the Lloyd codebook, cf. Section \ref{sec:conventionalscheme}, or the GMM codebook, cf. Section \ref{sec:proposedscheme}, contains quantized information regarding the spatial directions of each \ac{MT}'s channel and power loadings per stream.
We can extract the directional information of each codebook entry by simply performing an \ac{SVD} of each transmit covariance matrix, that is,
\begin{equation} \label{eq:extracted_dir}
    \mbQ_k = \mbX_{k} \mbT_k \mbX_{k}^\herm,
\end{equation}
where $\mbT_k$ contains the singular values in descending order, and taking the matrix $\bar{\mbX}_k$ which collects the first $\Nrx$ vectors of $\mbX_{k}$, as the respective directional information.
Accordingly, we have $\mc{Q} = \{\bar{\mbX}_1, \bar{\mbX}_2, \cdots, \bar{\mbX}_{K}\}$, which thus constitutes a directional codebook.
Note that, we have to ensure to take a codebook at sufficiently large \ac{SNR} in order to guarantee that the matrices ${\mbQ}_{k}$ all have a rank larger than or equal to $\Nrx$.
In our simulations this was the case with the codebooks constructed for an \ac{SNR} of $\SI{25}{\dB}$.

The described subspace-based method can be used in combination with both, the Lloyd codebook, cf. Section \ref{sec:conventionalscheme}, and the proposed GMM-based codebook, cf. Section \ref{sec:proposedscheme}.
In case of the subspaces extracted from the Lloyd codebook, the feedback per \ac{MT} is calculated after channel estimation by evaluating
\begin{equation}\label{eq:codebook_index_selection_dir_j}
    k^\star_j = \argmax_{k } ~\log_2 \det\left( \mathbf{I} + \frac{\rho}{\sigma_n^2 \Nrx} \hat{\mbH}_j \bar{\mbX}_k \bar{\mbX}_k^\herm \hat{\mbH}_j^\herm\right).
\end{equation}
When using the \ac{GMM}-based approach each \ac{MT} determines its feedback by simply evaluating the responsibilities:
\begin{equation} \label{eq:ecsi_index_j}
    k^\star_j = \argmax_{k } ~{p(k \mid \mby_j)}.
\end{equation}
In both cases, each \ac{MT} reports $k^\star_j$ to the \ac{BS} which represents each \ac{MT}'s channel using the subspace information associated with the respective codebook entry $\widetilde{\mbH}_j = \bar{\mbX}^\herm_{k^\star_j}$.
The \ac{BS} can then employ \ac{RBD} with the uniform power allocation policy, \ac{RCI}, or the iterative \ac{WMMSE} algorithm \cite[Algorithm~1]{HuCaYuQiXuYuDi21}, in order to jointly design the precoders $\mbM_j$ of all \acp{MT} $j \in \mathcal{J}$.

\subsection{Proposed Generative Modeling-based Method} \label{sec:gen_modeling}

As an alternative to the aforementioned subspace-based method, the channel matrix of each \ac{MT} may be modeled as a random variable, similar as in \cite{RaBoLu13, RaSaLu16}, and the average/ergodic achievable rate $R_j = \expec[R_j^{\text{inst}}]$ of each \ac{MT} $j\in\mathcal{J}$ is considered (note that the expectation is taken with respect to the channel distribution).
Then, the ergodic sum-rate maximization problem can be written as \cite{RaBoLu13, RaSaLu16}
\begin{equation} \label{eq:ergodic_sumrate}
    \max_{\{\mbM_j | j \in \mathcal{J}\}} \sum_{j=1}^{J} \expec[R_j^\text{inst}] ~~\text{s.t.} ~~\operatorname{tr}\left(\sum\nolimits_{j=1}^J\mbM_j \mbM_j^\herm\right) = \rho.
\end{equation}
In \cite{RaBoLu13, RaSaLu16}, the authors proposed the  \ac{SWMMSE} algorithm. It was shown that the algorithm is guaranteed to converge to the set of stationary points of the stochastic sum-rate maximization problem almost surely \cite{RaBoLu13, RaSaLu16}.
In each iteration step, the \ac{SWMMSE} algorithm requires channel samples that represent statistical information about each \ac{MT}.

The discussed conventional methods are unable to perform the \ac{SWMMSE} iterations since they are lacking of a generative model, i.e., a model that learns the underlying \ac{PDF} of the channels and allows to generate new samples that resemble the original channel distribution.
In contrast, the proposed \ac{GMM} approach is able to generate samples following the channel's distribution due to the \ac{GMM}'s sample generation ability.
This allows to jointly design the precoders via the \ac{SWMMSE} algorithm by exploiting statistical information about the \acp{MT} given their feedback information.
In particular, given the feedback index $k_j^\star$ of each \ac{MT}, see \eqref{eq:ecsi_index_j}, one can draw samples from the respective \ac{GMM} components via
\begin{equation} \label{eq:sampling_gmm}
    \mbh_{j,\text{sample}} \sim \calN_{\C}(\mbmu_{k^\star_j}, \mbC_{k^\star_j}),
\end{equation}
which represents statistical information about the channel of \ac{MT} $j$.
The \ac{BS} can subsequently design the precoders exploiting the \ac{SWMMSE} algorithm based on the generated samples utilizing the \ac{GMM}.
In order to feed the channel samples to the \ac{SWMMSE} algorithm, the channels have to be reshaped $\mbH_{j,\text{sample}} = \operatorname{unvec}(\mbh_{j,\text{sample}})$, cf., \cite{RaBoLu13, RaSaLu16} for more details on the \ac{SWMMSE}.
A summary of the proposed generative modeling-based precoder design method is given in Algorithm~\ref{alg:swmmse}.

\begin{algorithm}[t]
\small
\captionsetup{font=small}
\caption{Generative Modeling-based \ac{MU-MIMO} Precoder Design}
\label{alg:swmmse}
    \begin{algorithmic}[1]
    \STATE Set $i=0$, set max. iteration number $I_{\max}$, and randomly initialize the precoders such that $\operatorname{tr}(\sum_{j=1}^J\mbM_j \mbM_j^\herm) = \rho$.
    \REPEAT
    \STATE $\mbh^i_{j} \sim \calN_{\C}(\mbmu_{k^\star_j}, \mbC_{k^\star_j}), \forall j$
    \COMMENT{generate sample of the respective GMM component for each \ac{MT}}
    \STATE $\mbH^i_{j} \gets \operatorname{unvec}(\mbh^i_{j}),\forall j$
    \STATE $\mbU_j \gets \big(\sum_{m=1}^J \mbH^i_{j} \mbM_m \mbM_m^\herm \mbH^{i,\herm}_{j} + \sigma_j^2\mathbf{I} \big)^{-1} \mbH^i_{j} \mbM_j, \ \forall j$
    \STATE $ \mbW_j \gets \big(\mathbf{I} - \mbU_j^\herm \mbH^i_{j} \mbM_j  \big)^{-1}, \ \forall j $
    \STATE $ \mbZ_j \gets \mbM_j, \ \forall j $
    \STATE $ \mbA_j \gets \mbA_j + \beta \mathbf{I} + \sum_{m=1}^J \mbH^{i,\herm}_{j} \mbU_m \mbW_m \mbU_m^\herm \mbH^i_{j}, \ \forall j$
    \STATE $ \mbB_j \gets \mbB_j + \beta \mbZ_j + \mbH^{i,\herm}_{j} \mbU_j \mbW_j, \ \forall j$
    \STATE $ \mbM_j \gets \big( \mbA_j + \mu^\star \mathbf{I}  \big)^{-1} \mbB_j, \ \forall j$
    \STATE $i \gets i+1$
    \UNTIL{convergence or $i \geq I_{\max}$}
    \end{algorithmic}
\end{algorithm}

\section{Discussion on the Versatility of the Proposed Feedback Scheme}
\label{sec:discussion_universality}

\begin{figure}[t]
    \centering
    \includegraphics[scale=1.0]{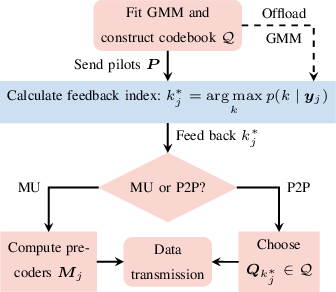}
    \caption{Flowchart of the proposed versatile feedback scheme. Red (blue) colored nodes are processed at the \ac{BS} (\acp{MT}).}
    \label{fig:flowchart}
\end{figure}

As discussed earlier, after fitting the \ac{GMM} centrally at the \ac{BS}, codebooks to support the point-to-point transmission mode can be constructed, cf. Section \ref{sec:proposedscheme}.
Then, the \ac{GMM} of the channels, cf. \eqref{eq:gmm_of_h}, is offloaded to every \ac{MT} within the coverage area of the \ac{BS}.
In the online phase, the \ac{BS} regularly sends pilots to the \acp{MT}.
Depending on the \ac{SNR} and the pilots, the \ac{GMM} of the observations, cf. \eqref{eq:gmm_y}, can be straightforwardly constructed from the offloaded \ac{GMM} of the channels.
With the help of the \ac{GMM} of the observations, the received observations at each \ac{MT} are processed to a feedback index $k_j^\star$ by evaluating the responsibility via \eqref{eq:ecsi_index_j}. 
Given the feedback information of each \ac{MT} at the \ac{BS}, the \ac{BS} can decide for the point-to-point or multi-user mode.
In case of the point-to-point mode, the \ac{BS} can simply select the codebook entry $\mbQ_{k^\star_j}$, cf. \eqref{eq:gmmcb_stage_2}, associated with \ac{MT} $j$ that should be served for data transmission, and no further processing is required.
In the multi-user mode, the \ac{BS} can represent each \ac{MT}'s channel using the subspace information associated with the respective $\bar{\mbX}^\herm_{k^\star_j}$ extracted from the high \ac{SNR} codebook, cf. \eqref{eq:extracted_dir}, and jointly design the precoders for multi-user transmission using either non-iterative (\ac{RBD} or \ac{RCI}) or iterative approaches (\ac{WMMSE}), thereby influencing the required processing time for designing the precoders.
Alternatively, the \ac{BS} can exploit the generative modeling capability of the \ac{GMM} and design the precoders using the \ac{SWMMSE} via sampling, cf. \eqref{eq:sampling_gmm}.
The proposed versatile feedback scheme is summarized as a flowchart in \Cref{fig:flowchart}, where red (blue) colored nodes represent processing steps that are performed at the \ac{BS} (\acp{MT}).

This flexibility is not provided by the discussed state-of-the-art approaches, since the feedback for the point-to-point mode associated with the current \ac{SNR} is determined by selecting an element via \eqref{eq:codebook_index_selection} out of a codebook constructed with this particular \ac{SNR} (cf. Section \ref{sec:conventionalscheme}), whereas in the multi-user case, the codebook entry selection is based on another codebook, i.e., the random quantization codebook, cf. \eqref{eq:codebook_index_selection_j}, or the directional codebook, cf. \eqref{eq:codebook_index_selection_dir_j}.

\section{Baseline Channel Estimators} 
\label{sec:baseline_channel_estimators}

Conventionally, the \ac{DL} channel is estimated firstly at each \ac{MT} and subsequently, the best fitting codebook entry is determined.
Thus, in this section, we present the baseline channel estimators which we consider in our simulations.
Since channel estimation takes place at each \ac{MT} separately, we present the estimators from a \ac{MT} perspective and drop the index $j$ in the following for brevity.
The \ac{MSE}-optimal channel estimate for the model~\eqref{eq:noisy_obs_p2p} is given by the \ac{CME} \( \expec[\mbh \mid \mby] \), cf., e.g.,~\cite[Section 8.1]{Sc91}.
However, the true channel \ac{PDF} is generally not known and, therefore, the \ac{CME} can generally not be calculated analytically.
Even if the true channel \ac{PDF} was known, the \ac{CME} \( \expec[\mbh \mid \mby] \) might still not have an analytic expression.

In this work, we use the recently proposed \ac{GMM}-based channel estimator \( \hhat_{\text{GMM}} \), see \eqref{eq:gmm_estimator_closed_form}, from \cite{KoFeTuUt21J} as a baseline.
The \ac{GMM}-based channel estimator is proven to asymptotically converge to the optimal \ac{CME} as the number of \ac{GMM} components $K$ is increased, with the restriction that $\mbA$ is invertible, see \eqref{eq:noisy_obs_p2p}.
In our case, we would have to fulfil that $n_p=\Ntx$.
However, even if $\mbA$ is not invertible ($n_p<\Ntx$) and for a moderate number for $K$, the \ac{GMM}-based channel estimator is a powerful estimator as shown in \cite{KoFeTuUt21J}.
The \ac{GMM}-based channel estimator utilizes the same \ac{GMM} as found in Subsection \ref{sec:proposedscheme}. 
In particular, the \ac{MT} can use the \ac{GMM} (obtained through offloading) to estimate the channel by evaluating:
\begin{equation}\label{eq:gmm_estimator_closed_form}
    \hhat_{\text{GMM}}(\mby) = \sum_{k=1}^K p(k \mid \mby) \hhat_{\text{LMMSE},k}(\mby)
\end{equation}
with the responsibilities $p(k \mid \mby)$ from \eqref{eq:responsibilities} and
\begin{equation}\label{eq:lmmse_formula}
    \hhat_{\text{LMMSE},k}(\mby) =
    \covhk \mbA^\herm (\mbA \covhk \mbA^\herm + \mbSigma)^{-1} (\mby - \mbA \meanhk) + \meanhk.
\end{equation}
Accordingly, the estimator \( \hhat_{\text{GMM}} \) is given by a weighted sum of \( K \) \ac{LMMSE} estimators---one for each component.
The weights \( p(k \mid \mby) \) are the probabilities that the current observation \( \mby \) corresponds to the \(k\)th component.

Another baseline is the \ac{LMMSE} estimator, which utilizes the sample covariance matrix $\mbC_s = \frac{1}{M} \sum_{m=1}^M \mbh_m \mbh_m^\herm$, which is calculated using the same set of training samples to fit the \ac{GMM}, and calculate channel estimates as (cf. \cite{fesl2021centralized, KoFeTuUt21J}):
\begin{equation}\label{eq:sample_cov}
    \hhat_{\text{LMMSE}} = \mbC_s \mbA^\herm (\mbA \mbC_s \mbA^\herm + \mbSigma)^{-1} \mby.
\end{equation}

Lastly, compressive sensing approaches commonly assume that the channel exhibits a certain structure: $\mbh \approx \mbD \mbt$, where $\mbD = \mbD_{\text{rx}} \otimes (\mbD_{\text{tx,h}} \otimes \mbD_{\text{tx,v}})$ is a \textit{dictionary} with oversampled \ac{DFT} matrices $\mbD_{\text{rx}}$, $\mbD_{\text{tx,h}}$, and $\mbD_{\text{tx,v}}$ (cf., e.g., \cite{AlLeHe15}), because we have a \ac{URA} at the \ac{BS} and a \ac{ULA} at the \ac{MT}.
A compressive sensing algorithm like \ac{OMP}~\cite{Gharavi} can now be used to obtain a sparse vector \( \mbt \), and the estimated channel is then given by
\begin{equation} \label{eq:omp_est}
    \hhat_{\text{OMP}} = \mbD \mbt.
\end{equation}
Since the sparsity order is not known, but the algorithm's performance crucially depends on it, we use a genie-aided approach to obtain a bound on the performance of the algorithm. Specifically, we use the true channel (perfect \ac{CSI} knowledge) to choose the optimal sparsity order.

\section{Complexity Analysis} \label{sec:comp_ana}

The responsibilities in \eqref{eq:responsibilities} are calculated by evaluating Gaussian densities.
Since the \ac{GMM}'s covariance matrices and mean vectors do not change for different observations, the inverse and the determinant of the densities can be pre-computed.
Therefore, the online evaluation of the responsibilities \( p(k \mid \mby) \) in~\eqref{eq:responsibilities} is dominated by matrix-vector multiplications and has a complexity of \( \calO(L^2) \) per \ac{GMM} component, with $L=\Nrx n_p$.

Correspondingly, determining the feedback using the \ac{GMM} via \eqref{eq:ecsi_index} has a complexity of \( \calO(K \Nrx^2 n_p^2) \).
Recall that in this case, no channel estimation needs to be conducted.
One huge advantage is, that the complexity does not scale with the number of transmit antennas $\Ntx$.
This is particularly beneficial if the \ac{BS} is equipped with many antennas, as it is the case for massive \ac{MIMO} systems.
Moreover, the proposed method allows for parallelization with respect to the number of components $K$, i.e., all of the $K$ responsibilities can be evaluated in parallel.

When using the conventional approach of first estimating the channel and then searching for the best codebook entry, the complexity depends on the channel estimation complexity and the complexity of the selection method from \eqref{eq:codebook_index_selection}.
Among all considered conventional approaches, the \ac{GMM} estimator from \eqref{eq:gmm_estimator_closed_form} in combination with the selection method that maximizes the rate expression from \eqref{eq:codebook_index_selection} performed best in our simulations, cf. Section \ref{sec:sim_results}.
Evaluating the \ac{GMM} estimator from \eqref{eq:gmm_estimator_closed_form} has a complexity of 
\(\calO(K\Nrx^2 n_p^2 + K\Nrx^2\Ntx n_p)\), due to the calculation of the responsibilities \( p(k \mid \mby) \) and the evaluation of the \ac{LMMSE} filters from \eqref{eq:lmmse_formula} \cite{KoFeTuUt21J}.
The responsibilities \( p(k \mid \mby) \) from~\eqref{eq:responsibilities} to determine the feedback using the \ac{GMM} according to \eqref{eq:ecsi_index} are the same responsibilities which are needed to evaluate the \ac{GMM} estimator from \eqref{eq:gmm_estimator_closed_form}. 
Evaluating the \ac{GMM} estimator further requires the calculation of the \ac{LMMSE} filters from~\eqref{eq:lmmse_formula}.
Thus, in terms of \ac{FLOPS} our proposed method from \eqref{eq:ecsi_index} is in any case of lower complexity for the \ac{MT} as compared to evaluating the \ac{GMM} estimator from~\eqref{eq:gmm_estimator_closed_form}. 
In addition to that, with the conventional approach, the estimated channel has to be further processed to a feedback index by evaluating \eqref{eq:codebook_index_selection}.
The complexity of this selection method is \(\calO(K \Ntx \Nrx^2 + K \Nrx^3)\), when exploiting the QR decomposition \cite[Section~5.2]{Go89}.

This complexity analysis also holds for the multi-user case since the feedback is determined at each \ac{MT} separately by conducting similar steps, i.e., by first estimating the channel and then evaluating either \eqref{eq:codebook_index_selection_j} in the case of random codebooks, or \eqref{eq:codebook_index_selection_dir_j} in the case of the directional codebook.

\emph{Kronecker Approximation for Reducing the Offloading Amount:} In order for a \ac{MT} to be able to compute feedback indices, the parameters of the \ac{GMM} \( f_{\mbh}^{(K)} \) need to be offloaded to the \ac{MT} upon entering the \ac{BS}' coverage area.
As demonstrated in a numerical example in Table~\ref{tab:num_params}, the number of \ac{GMM} parameters can be quite large.
This is mainly due to the large number of parameters of the \ac{GMM}'s covariance matrices.
In order to reduce the number of \ac{GMM} parameters, we can incorporate model-based insights without influencing the online computational complexity.
For spatial correlation scenarios, a well-known assumption is that the scattering in the vicinity of the transmitter and receiver are independent of each other, cf.~\cite{KeScPeMoFr02}.
Similarly, as in \cite{KoFeTuUt21J}, we use this assumption to constrain the \ac{GMM} covariance matrices to a Kronecker factorization with fewer parameters, i.e., we construct a \ac{GMM} consisting of covariance matrices of the form \( \covhk = \mbC_{\text{tx},k} \otimes \mbC_{\text{rx},k} \).
Thus, instead of fitting a single unconstrained \ac{GMM} with $N\times N$-dimensional covariances, a transmit-side (receive-side) \ac{GMM} with $N_{\text{tx}} \times N_{\text{tx}}$ ($N_{\text{rx}} \times N_{\text{rx}}$)-dimensional covariances and $K_{\text{tx}}$ ($K_{\text{rx}}$) components is fitted.
Thereafter, a \( K = \Ktx\Krx \)-components \ac{GMM} with $N\times N$-dimensional covariances is obtained by combinatorially computing all Kronecker products of the respective transmit- and receive-side covariance matrices.
It was observed in \cite{KoFeTuUt21J} that the Kronecker \ac{GMM} performs almost equally well as compared to the unconstrained \ac{GMM}.
The advantages of the Kronecker \ac{GMM} are a lower offline training complexity, the ability to parallelize the fitting process, and the need for fewer training samples since the Kronecker \ac{GMM} has much fewer parameters.

\begin{table}[t]
\renewcommand{\arraystretch}{1.1}
    \begin{center}
    \resizebox{\columnwidth}{!}{
    \renewcommand{\arraystretch}{2}
    \begin{tabular}{|l|c|c|c|}
    \hline
          \textbf{Name}            & \textbf{Online Complexity}                  & \textbf{Covariance Parameters}                                                      & \begin{tabular}{@{}l@{}} $(\Ntx, \Nrx) = (32,16)$, \\ $(\Ktx, \Krx) = (16,4)$\end{tabular}\\ \hline
    Full & \( \calO(K \Nrx^2 n_p^2) \) & \( \frac{1}{2} K N(N+1) \)                                        & \( 8.4 \cdot 10^6 \)                                                              \\ \hline
    Kronecker & \( \calO(K \Nrx^2 n_p^2) \) & $ \frac{1}{2}\Krx \Nrx(\Nrx+1)$  $+ \frac{1}{2}\Ktx \Ntx(\Ntx+1)$ & \( 9 \cdot 10^3 \)                                                              \\ \hline
    \end{tabular}
    }
    \end{center}
\caption{Analysis of the number of parameters of the (structured) \ac{GMM}.}
\label{tab:num_params}
\end{table}

Table \ref{tab:num_params} illustrates exemplarily the difference in the number of \ac{GMM} covariance parameters (taking symmetries into account), where we plug in the simulation parameters of one of the settings with $B=6$, which we consider in Section~\ref{sec:sim_results}.
We can see that, with the Kronecker \ac{GMM}, the number of parameters that need to be offloaded is drastically reduced.
For the remaining settings considered in Section~\ref{sec:sim_results}, the reduction factors due to the Kronecker \ac{GMM} are in the order of approximately $10^2$ to $10^3$.
For this reason, we solely consider the Kronecker \ac{GMM} in Section~\ref{sec:sim_results}.

\section{Simulation Results} \label{sec:sim_results}

With trends towards massive \ac{MIMO}, both the \ac{BS} and the \acp{MT} are equipped with many antennas~\cite{AnBuChHaLoSoZh14}.
The \ac{BS} equipped with a \ac{URA} has in total $\Ntx=\Ntxh \Ntxv$ antenna elements, with $\Ntxh$ horizontal and $\Ntxv$ vertical elements.
At the \ac{MT}, we have a \ac{ULA} with $\Nrx$ elements.
We consider $B$ feedback bits and thus $K = 2^B$.
We generate datasets with $30 \cdot 10^3$ channels for both the \ac{UL} and \ac{DL} domain of the scenario described in Section \ref{sec:data_generation}: $\mathcal{H}^{\text{UL}}$ and $ \mathcal{H}^{\text{DL}} $.
The data samples are normalized such that \( \expec[\|\mbh\|^2] = N = \Ntx \Nrx \) holds for the vectorized channels. 
We further set $\rho=1$, which allows us to define the \ac{SNR} as \( \frac{1}{\sigma_n^2} \) for all \acp{MT}, i.e., when $\sigma^2_j = \sigma^2_n, \forall j \in \mathcal{J}$.
We split the two sets $\mathcal{H}^{\text{UL}}$ and $\mathcal{H}^{\text{DL}}$ into a training set with $M = 20 \cdot 10^3$ samples, and the remaining samples constitute an evaluation set, viz., $\mathcal{H}_{\text{train}}^{\text{UL}}, \mathcal{H}_{\text{eval}}^{\text{UL}}, \mathcal{H}_{\text{train}}^{\text{DL}}, \ \text{and} \ \mathcal{H}_{\text{eval}}^{\text{DL}}.$
The following transmit strategies are always evaluated on \( \mathcal{H}_{\text{eval}}^{\text{DL}} \), i.e., in the \ac{DL} domain.
When we fit the \ac{GMM} based on \( \mathcal{H}_{\text{train}}^{\text{UL}} \), we transpose all elements of the set to emulate a \ac{DL}, cf. \cite{TuKoBaXuUt21, fesl2021centralized, TuKoRiFeBaXuUt21}.

\subsection{Point-to-point MIMO}

\begin{figure}[tb]
    \centering
    \includegraphics[scale=1.0]{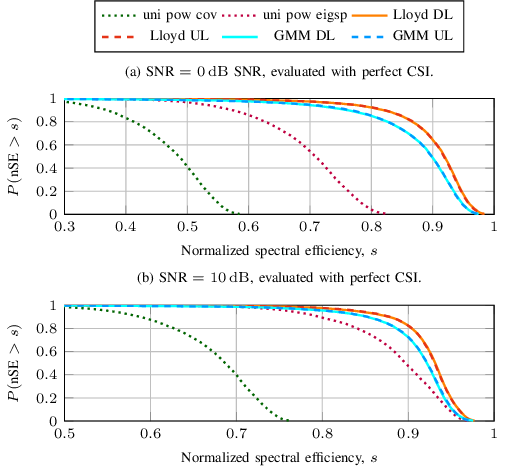}
    \caption{Empirical \acp{cCDF} of the normalized (by the optimal transmit strategy) spectral efficiencies achieved with different codebooks and transmit strategies evaluated with perfect \ac{CSI}, for a system with $\Ntx=32$, $\Nrx=16$, and $B=6$ bits.}
    \label{fig:cbatSNR0and10}
\end{figure}

In the single-user case, we depict the \ac{nSE} as performance measure. 
The spectral efficiency achieved with a given transmit covariance matrix is normalized by the spectral efficiency achieved with the optimal transmit covariance matrix, which is given by decomposing the channel into $\Nrx$ parallel streams and employing water-filling \cite{Telatar99capacityof}.
The empirical \ac{cCDF} $P(\text{nSE}>s)$ of the normalized spectral efficiency, corresponds to the empirical probability that \ac{nSE} exceeds a specific value $s$.

We consider the following baseline transmit strategies:
The curves labeled ``{uni pow cov}'' represent uniform power allocation where the transmit covariance matrix is given by \( \mbQ = \frac{\rho}{\Ntx} \mathbf{I} \). 
In this case, no \ac{CSI} knowledge or codebook is used.
Moreover, ``{uni pow eigsp}'' corresponds to the transmit strategy where a transmit covariance matrix is calculated by allocating equal power on the eigenvectors of the channel.
That is, the channel is decomposed into $\Nrx$ parallel streams and $\frac{\rho}{\Nrx}$ power is allocated to each stream.
Note that this approach is infeasible because the \ac{BS} would require full knowledge of the \ac{DL} channel (or its eigenvectors).

\begin{figure}[tb]
    \centering
    \includegraphics[scale=1.0]{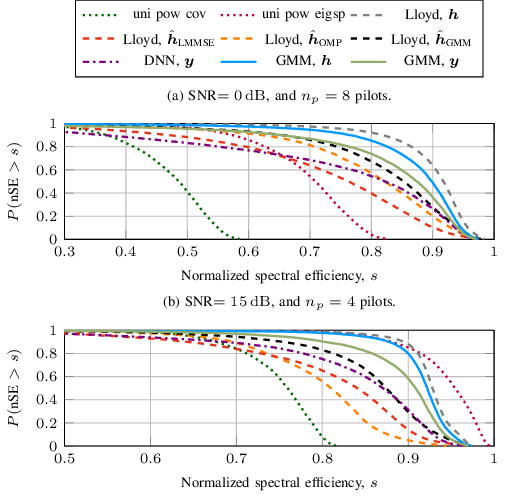}
    \caption{Empirical \acp{cCDF} of the normalized (by the optimal transmit strategy) spectral efficiencies achieved with different codebooks and transmit strategies evaluated for a system with $\Ntx=32$, $\Nrx=16$, different \acp{SNR}, different number of pilots $n_p$, and $B=6$ bits.}
    \label{fig:cbatSNR0P8_SNR15P4}
\end{figure}

In the following, the simulation parameters are $\Ntx=32$ ($\Ntxh=8, \Ntxv=4$), $\Nrx=16$ and $B=6$ bits, thus $K=64$ ($K_\text{tx}=16$, $K_\text{rx}=4$).
In \Cref{fig:cbatSNR0and10}(a), we set the $\text{SNR}=\SI{0}{dB}$.
The conventional codebook construction approach (cf. Section \ref{sec:conventionalscheme}) is denoted by ``{Lloyd UL/DL}'', depending on whether $\mathcal{H}_{\text{train}}^{\text{UL}}$ or $\mathcal{H}_{\text{train}}^{\text{DL}}$ is used as training data to construct the codebooks, respectively.
With these approaches, the codebook is known to the \ac{BS} and the \ac{MT} and, additionally, perfect \ac{CSI} is assumed at the \ac{MT}.
Each \ac{MT} then selects the best possible codebook entry by evaluating~\eqref{eq:codebook_index_selection}.
We can observe that, using \ac{DL} or \ac{UL} training data, results in approximately the same performance.
The proposed codebook construction and encoding scheme is denoted by ``{GMM UL/DL}'', where we either use $\mathcal{H}_{\text{train}}^{\text{UL}}$ or $\mathcal{H}_{\text{train}}^{\text{DL}}$ as training data to fit the \ac{GMM} and to construct the codebook as described in Section \ref{sec:proposedscheme}.
With our proposed approach, the knowledge of the codebook at the \ac{MT} is not required.
After offloading the \ac{GMM} to the \ac{MT} and given perfect \ac{CSI} knowledge, the \ac{MT} can then determine the feedback index by evaluating \eqref{eq:pcsi_index}.
Again, using \ac{DL} or \ac{UL} training data results in approximately the same performance, which is in accordance with the findings from \cite{TuKoBaXuUt21, fesl2021centralized, TuKoRiFeBaXuUt21}.
The proposed \ac{GMM} approach performs slightly worse than the Lloyd clustering approach, which is a consequence of the perfect \ac{CSI} assumption.
In \Cref{fig:cbatSNR0and10}(b), we set $\text{SNR}=\SI{10}{dB}$, and observe similar results.

However, assuming perfect \ac{CSI} at the \ac{MT} is not feasible.
In the following, we consider imperfect \ac{CSI}, i.e., systems with reduced pilot overhead ($n_p \leq \Ntx$).
In the remainder, we consider \ac{UL} training data exclusively.
Thus, we omit writing ``UL'' in the legend from now on.

\begin{figure}[tb]
    \centering
    \includegraphics[scale=1.0]{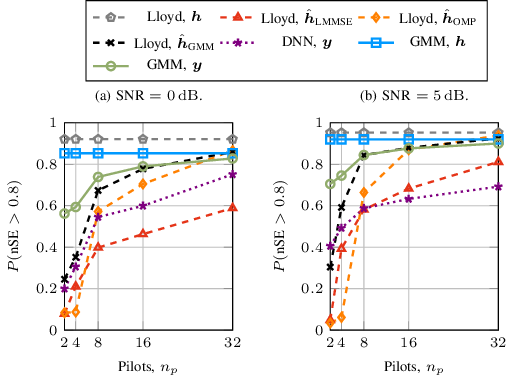}
    \caption{The probability that the \ac{nSE} of a certain transmit strategy exceeds $s=80\%$ of the optimal transmit strategy's spectral efficiency for a varying number of pilots,  for a system with $\Ntx=32$, $\Nrx=16$, and $B=6$ bits.}
    \label{fig:ccdf_overpilots_32x16_snr0_snr5}
\end{figure}

In \Cref{fig:cbatSNR0P8_SNR15P4}(a), the $\text{SNR}=\SI{0}{dB}$ and we have $n_p=8$.
We depict results for the conventional Lloyd codebook construction approach (cf. Section \ref{sec:conventionalscheme}), where we first estimate the channel either via \ac{OMP}~\eqref{eq:omp_est}, the \ac{LMMSE} approach~\eqref{eq:sample_cov}, or via the \ac{GMM} estimator~\eqref{eq:gmm_estimator_closed_form}, and then select a transmit covariance matrix by evaluating~\eqref{eq:codebook_index_selection} given the estimated channel: ``{{Lloyd}, $\hhat_{\text{OMP}}$}'', ``{{Lloyd}, $\hhat_{\text{LMMSE}}$}'', or ``{{Lloyd}, $\hhat_{\text{GMM}}$}'', respectively.
Moreover, we compare to the \ac{SNR}-independent \ac{DNN} approach from \cite{TuKoBaXuUt21}, denoted by ``DNN $\mby$'',  where a classifier is employed to directly map the observation to a feedback index that specifies an element from the Lloyd codebook.
During the training phase, the \ac{DNN} was provided with input-output pairs $\{(\mbY_m=\uvect(\mby_m), k^\star_m)\}_{m=1}^M$ for an \ac{SNR} range of $\SI{0}{dB}$ to $\SI{25}{dB}$, with $\SI{5}{dB}$ steps.
We employ random search \cite{BeBe12} to determine the hyperparameters of the \ac{DNN}.
The \ac{DNN} consists of $D_\text{CM}$ convolutional modules, which comprise a convolutional layer, a batch normalization, and an activation function, where $D_\text{CM}$ is randomly chosen from the range $[2, 5]$.
Each of the convolutional layers consists of $D_\text{K}$ kernels, where $D_\text{K}$ is randomly chosen within $[16, 64]$.
After a subsequent two-dimensional max-pooling, the features are flattened, and a fully connected layer is employed with an output dimension of~$K$.
Depending on the randomly drawn parameters, the number of \ac{DNN} parameters is at least the same as or even higher than the number of \ac{GMM} parameters, and increases with the number of pilots.
Moreover, note that a separate \ac{DNN} per pilot configuration is needed.
The complexity of the \ac{DNN} approach is $\mathcal{O}(K D_\text{K} \Nrx n_p)$.
Further details can be found in \cite{TuKoBaXuUt21}.
As can be seen, estimating the channel via the \ac{GMM} estimator gives the best performance when considering the conventional channel estimation-based approaches. 
The \ac{DNN} approach, which also does not require any codebook knowledge, similar to our proposed \ac{GMM}-based feedback scheme, achieves comparable performance as ``{{Lloyd}, $\hhat_{\text{OMP}}$}'' or ``{{Lloyd}, $\hhat_{\text{LMMSE}}$}''.
In contrast, with the proposed approach (cf. Section \ref{sec:proposedscheme}) denoted by ``{{GMM}, $\mby$}'', where we bypass channel estimation and directly evaluate~\eqref{eq:ecsi_index} for determining a feedback index, we achieve even better performance as compared to any of the conventional approaches.
With the curves ``{{Lloyd}, $\mbh$}'' and ``{{GMM}, $\mbh$}'', we depict the results for the utopian case of assuming perfect \ac{CSI} knowledge at the \ac{MT}.
Although the Lloyd approach performs well if perfect \ac{CSI} is available at the \ac{MT}, the performance suffers significantly from \ac{CSI} imperfections (due to noise and low pilot overhead).
In contrast, the proposed \ac{GMM}-based feedback scheme is superior in case of imperfect \ac{CSI} available at the \ac{MT}, which resembles practical system deployments.
A similar observation can also be made in \Cref{fig:cbatSNR0P8_SNR15P4}(b), where the $\text{SNR}=\SI{15}{dB}$ and we only have $n_p=4$ pilots.

\begin{figure}[tb]
    \centering
    \includegraphics[scale=1.0]{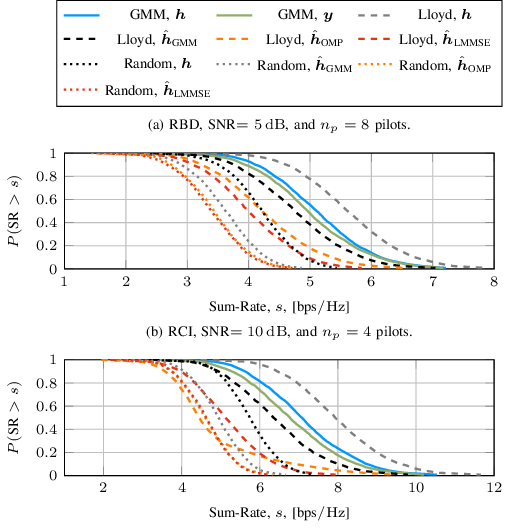}
    \caption{Empirical \acp{cCDF} of the sum-rate ($\Ntx=16$, $\Nrx=4$, and $J=4$ \acp{MT}) achieved with different feedback approaches and precoding techniques for different \acp{SNR}, different number of pilots $n_p$, and $B=6$ bits.}
    \label{fig:MUcbatSNR5P8_SNR10P4_RBDnew}
\end{figure}

In \Cref{fig:ccdf_overpilots_32x16_snr0_snr5}(a), we set $\text{SNR}=\SI{0}{dB}$ and in \Cref{fig:ccdf_overpilots_32x16_snr0_snr5}(b), we have $\text{SNR}=\SI{5}{dB}$, where we fix $s=0.8$, thus, we consider $P(\text{nSE}>0.8)$ for a varying number of pilots $n_p$.
We see that our proposed low-complexity feedback scheme is beneficial in the low number of pilots regime and outperforms the conventional approaches, which either require both channel estimation and the feedback evaluation via~\eqref{eq:codebook_index_selection} or the \ac{DNN}-based approach.

\subsection{Multi-user MIMO}

In this subsection, we present simulation results for the multi-user setup and depict the sum-rate as performance measure, which is given by $\sum_{j=1}^J R_j^{\text{inst}}$, cf. \eqref{eq:inst_sumrate}.
We depict the results for $2{,}500$ constellations, where for each constellation, we draw $J$ \acp{MT} randomly from our evaluation set \( \mathcal{H}_{\text{eval}}^{\text{DL}} \).
The empirical \ac{cCDF} $P(\text{SR}>s)$ of the sum-rate, is used to depict the empirical probability that the \ac{SR} exceeds a specific value $s$.

In the following, with ``GMM, $\mbh$'' and ``GMM, $\mby$'' we denote the cases, where either perfect \ac{CSI} $\mbh_j$ is assumed or the observations $\mby_j$ are used at each \ac{MT} $j$ to determine a feedback index using the \ac{GMM} feedback encoding approach, cf. \eqref{eq:ecsi_index_j}.
We omit the index $j$ in the legend for notational convenience.
The channel of each \ac{MT} is then represented by the subspace information extracted from the high-\ac{SNR} \ac{GMM} codebook, cf. Section \ref{sec:sub_method}.
With ``Lloyd, $\mbh$'', ``Lloyd, $\hhat_{\text{GMM}}$'', ``Lloyd, $\hhat_{\text{OMP}}$'', and ``Lloyd, $\hhat_{\text{LMMSE}}$'',
or with ``Random, $\mbh$'', ``Random, $\hhat_{\text{GMM}}$'', ``Random, $\hhat_{\text{OMP}}$'', and ``Random, $\hhat_{\text{LMMSE}}$'',
we denote the cases where either perfect \ac{CSI} is assumed at each \ac{MT}, or the channel is firstly estimated at each \ac{MT} and then the index of the best fitting subspace entry of the high-\ac{SNR} Lloyd codebook or of the random codebook, is fed back from each \ac{MT} to the BS, cf. Sections \ref{sec:sub_method} and \ref{sec:conventional_method}.

The above mentioned approaches are evaluated using either \ac{RBD}, \ac{RCI}, or the iterative \ac{WMMSE} to jointly design the precoders $\mbM_j, \ \forall j \in \mathcal{J}$.
In case of \ac{RBD} and \ac{RCI}, the used regularization factor is $\frac{J\Nrx \sigma_n^2}{\rho}$, and the precoders are normalized to satisfy the transmit power constraint, cf. \cite{StHa08, PeHoSw05, LeLeHoLe08}.
In the case of the iterative \ac{WMMSE}, we use \cite[Algorithm~1]{HuCaYuQiXuYuDi21}.
Additionally, with ``GMM samples, $\mbh$'' and ``GMM samples, $\mby$'', we denote the cases where we generate samples which represent each \ac{MT}'s distribution using the \ac{GMM} and feed them to the SWMMSE algorithm, cf. Section~\ref{sec:gen_modeling}.
In all iterative approaches, we set $I_{\max}=300$ iterations.

\begin{figure}[tb]
    \centering
    \includegraphics[scale=1.0]{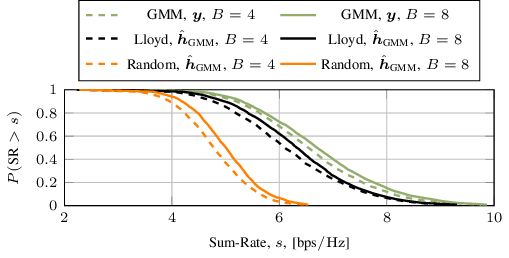}
    \caption{Empirical \acp{cCDF} of the sum-rate ($\Ntx=16$, $\Nrx=4$, and $J=4$ \acp{MT}) achieved with different feedback approaches when \ac{RBD} is employed for a system with an $\text{SNR}=\SI{10}{dB}$, $n_p=4$ pilots, and different bits $B$.}
    \label{fig:MUcbatSNR5P8_SNR10P4_RBDnew_K16_K256_imperfect}
\end{figure}

In the following, the simulation parameters are $\Ntx=16$ ($\Ntxh=4, \Ntxv=4$), $\Nrx=4$ and $B=6$ bits, thus $K=64$ ($K_\text{tx}=16$, $K_\text{rx}=4$).
Accordingly, we have $J=4$ \acp{MT}.
In \Cref{fig:MUcbatSNR5P8_SNR10P4_RBDnew}(a), the $\text{SNR}=\SI{5}{dB}$ and we have $n_p=8$ pilots.
In this case, we use \ac{RBD} in order to jointly design the precoders.
We can observe that the random codebook performs worst.
Even with perfect \ac{CSI} assumed at the \acp{MT}, the random codebook approach cannot compete with the environment-aware approaches.
The Lloyd directional codebook approach yields the best performance, if the \ac{GMM}-based channel estimator from \eqref{eq:gmm_estimator_closed_form} is used prior to codebook entry selection.
Similar to the point-to-point \ac{MIMO} case, we can observe that the chosen channel estimator significantly impacts the performance.
That is, using the \ac{LMMSE} estimator from \eqref{eq:sample_cov} or the genie \ac{OMP} from \eqref{eq:omp_est} yield worse results as compared to the \ac{GMM}-based channel estimator.
Furthermore, our proposed \ac{GMM}-based feedback approach (``GMM, $\mby$'') even outperforms the best conventional approach (``Lloyd, $\hhat_{\text{GMM}}$'').
A similar behavior can be observed in \Cref{fig:MUcbatSNR5P8_SNR10P4_RBDnew}(b), where we increased the $\text{SNR}$ to $\SI{10}{dB}$ and decreased the number of pilots $n_p=4$, and use \ac{RCI} in order to jointly design the precoders.

\begin{figure}[tb]
    \centering
    \includegraphics[scale=1.0]{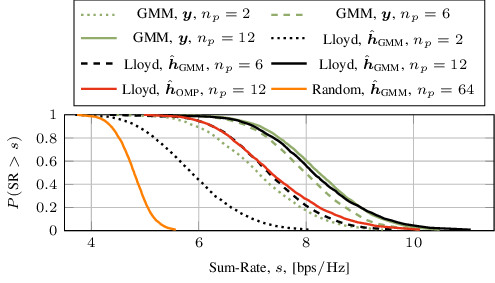}
    \caption{Empirical \acp{cCDF} of the sum-rate ($\Ntx=64$, $\Nrx=4$, and $J=16$ \acp{MT}) achieved with different feedback approaches when \ac{RBD} is employed for systems with an $\text{SNR}=\SI{5}{dB}$, different number of pilots $n_p$, and $B=6$ bits.}
    \label{fig:MUcbatSNRXPX_SNR10P4_RBDnew_K64_imperfect_lowp}
\end{figure}

In the following two figures (\Cref{fig:MUcbatSNR5P8_SNR10P4_RBDnew_K16_K256_imperfect} and \Cref{fig:MUcbatSNRXPX_SNR10P4_RBDnew_K64_imperfect_lowp}) we restrict our analysis to \ac{RBD} as the precoder design algorithm for sake of brevity.
The purpose of the next two figures is to quantify the performance gains obtained with our proposed approach from different perspectives.
In particular, in \Cref{fig:MUcbatSNR5P8_SNR10P4_RBDnew_K16_K256_imperfect} we have $\text{SNR}=\SI{10}{dB}$ and $n_p=4$, and we consider systems with $B=4$ bits, thus, $K=16$ ($K_\text{tx}=8$, $K_\text{rx}=2$), or $B=8$ bits, thus $K=256$ ($K_\text{tx}=32$, $K_\text{rx}=8$) and compare the performance of our proposed \ac{GMM}-based feedback approach (``GMM, $\mby, B\in\{4,8\}$'') to the best performing conventional Lloyd directional codebook (``Lloyd, $\hhat_{\text{GMM}}, B\in\{4,8\}$'') and random codebook (``Random, $\hhat_{\text{GMM}}, B\in\{4,8\}$'') approaches, which use the \ac{GMM}-based channel estimator in the channel estimation phase.
We can observe that our proposed feedback approach is superior to the conventional methods. 
In particular, our proposed feedback approach with only $B=4$ bits (``GMM, $\mby, B=4$'') even outperforms the conventional Lloyd directional codebook with twice as much, i.e., $B=8$, bits (``Lloyd, $\hhat_{\text{GMM}}, B=8$'').

In \Cref{fig:MUcbatSNRXPX_SNR10P4_RBDnew_K64_imperfect_lowp}, we consider a system with more transmit antennas and, accordingly, more \acp{MT}.
The simulation parameters are $\Ntx=64$ ($\Ntxh=8, \Ntxv=8$), $\Nrx=4$, and $B=6$ bits, thus, $K=64$ ($K_\text{tx}=16$, $K_\text{rx}=4$), and the $\text{SNR}=\SI{5}{dB}$. 
Accordingly, we have $J=16$ \acp{MT}.
This time, we depict the performances for a varying number of pilots $n_p$.
For a fixed number of pilots $n_p$, our proposed approach (``GMM, $\mby, n_p\in \{2,6,12\}$'') outperforms the conventional approaches (``Lloyd, $\hhat_{\text{GMM}}, n_p\in \{2,6,12\}$'' and ``Lloyd, $\hhat_{\text{OMP}}^{}, n_p=12$'').
Moreover, with only $n_p=2$ pilots, our proposed feedback approach (``GMM, $\mby, n_p=2$''), almost achieves the same performance as the conventional approaches, which require $n_p=6$ pilots in the case of ``Lloyd, $\hhat_{\text{GMM}}, n_p=6$'', or even $n_p=12$ in the case of ``Lloyd, $\hhat_{\text{OMP}}^{}, n_p=12$'' (i.e., the \acp{MT} are unaware of the \ac{GMM} and use the \ac{OMP} channel estimator).
If random codebooks are used, the performance with even a large pilot overhead, i.e. $n_p=64$ pilots, is poor (``Random, $\hhat_{\text{GMM}}, n_p=64$'').
Thus, with our proposed approach, systems with lower pilot overhead can be deployed, which would inherently increase the system throughput.
Additionally, with fewer pilots, the complexity of determining the feedback index at the \acp{MT} with the proposed \ac{GMM}-based approach decreases, cf. Section \ref{sec:comp_ana}.

So far, we have only considered non-iterative precoding algorithms, i.e., \ac{RBD} and \ac{RCI}. 
In the remainder, we will focus our analysis on the iterative \ac{WMMSE} and the \ac{SWMMSE} precoding techniques.
Due to the exclusive usage of channel directional information, i.e., no channel magnitude information is fed back to the \ac{BS}, as in \cite{RaJi08} (cf. Section \ref{sec:conventional_method}), in case of random codebooks, or the Lloyd directional codebook from Section \ref{sec:sub_method}, or the \ac{GMM} directional codebook from Section \ref{sec:sub_method}, changing the number of streams $d$ impacts the overall sum-rate which can be achieved using the iterative \ac{WMMSE}.
In fact, we observed that depending on the \ac{SNR}, the number of pilots $n_p$, the chosen channel estimator (for the conventional approaches), and the selected codebook, the performance can be improved by varying $d \in \{1,2, \cdots, \Nrx\}$, and then setting $d$ to the value which gives the best average performance.
In a practical deployment, the parameter $d$ can be pre-adjusted in the offline phase at the \ac{BS} by emulating a \ac{DL} system (for example using the set $\mathcal{H}_{\text{eval}}^{\text{UL}}$).
However, things are different with the \ac{SWMMSE}.
There we observed that setting $d=\Nrx$ always yields the best performance.
Intuitively, due to the sampling involved in the design procedure of the \ac{SWMMSE}, (average) channel magnitude information is provided to and exploited by the \ac{SWMMSE} algorithm, which enables the \ac{SWMMSE} to adjust the stream powers accordingly.

\begin{figure}[tb]
    \centering
    \includegraphics[scale=1.0]{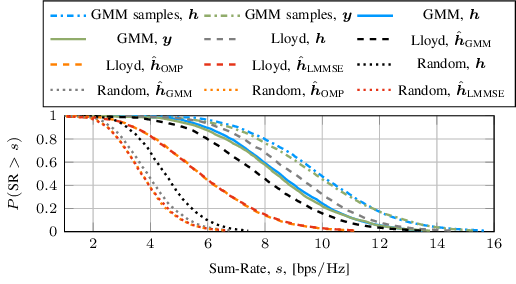}
    \caption{Empirical \acp{cCDF} of the sum-rate ($\Ntx=16$, $\Nrx=4$, and $J=4$ \acp{MT}) achieved with different feedback approaches when the iterative \ac{WMMSE} or the \ac{SWMMSE} are employed, for a system with an $\text{SNR}=\SI{5}{dB}$, $n_p=8$ pilots, and $B=6$ bits.}
    \label{fig:MUcbatSNR5P8_SNR10P4_WMMSEd1_new}
\end{figure}

In the remainder, the simulation parameters are again $\Ntx=16$ ($\Ntxh=4, \Ntxv=4$), $\Nrx=4$, yielding $J=4$ \acp{MT}, and $B=6$ bits, thus, $K=64$ ($K_\text{tx}=16$, $K_\text{rx}=4$).
In \Cref{fig:MUcbatSNR5P8_SNR10P4_WMMSEd1_new}, the $\text{SNR}=\SI{5}{dB}$ and we have $n_p=8$ pilots.
This is the same simulation setting as in \Cref{fig:MUcbatSNR5P8_SNR10P4_RBDnew}(a), where \ac{RBD} was used in order to design the precoders.
By comparing \Cref{fig:MUcbatSNR5P8_SNR10P4_WMMSEd1_new} and \Cref{fig:MUcbatSNR5P8_SNR10P4_RBDnew}(a), we can conclude that by using the iterative precoding techniques, the performances of the conventional and the proposed feedback approaches are improved tremendously. 
We can observe, that also in the case of iterative precoding techniques, the random codebook approach performs worst.
Also in this case, the Lloyd directional codebook approach yields the best performance, if the \ac{GMM}-based channel estimator from \eqref{eq:gmm_estimator_closed_form} is used prior to codebook entry selection whereas using the \ac{LMMSE} estimator from \eqref{eq:sample_cov} or the genie \ac{OMP} from \eqref{eq:omp_est} deteriorates the performance.
Furthermore, our proposed low-complexity \ac{GMM}-based feedback approach (``GMM, $\mby$'') again outperforms the best conventional approach (``Lloyd, $\hhat_{\text{GMM}}$'').
With our generative modeling-based approach from Section \ref{sec:gen_modeling}, i.e., the \ac{SWMMSE} with samples generated by the \ac{GMM}, denoted by ``GMM samples, $\mbh$'' or ``GMM samples, $\mby$'', we even outperform the performance bound of the Lloyd directional codebook approach (which uses the iterative \ac{WMMSE}) with perfect \ac{CSI} assumed at the \acp{MT} (``Lloyd, $\mbh$'').
This shows the great potential of the generative modeling ability of the \ac{GMM}.

\begin{figure}[tb]
    \centering
    \includegraphics[scale=1.0]{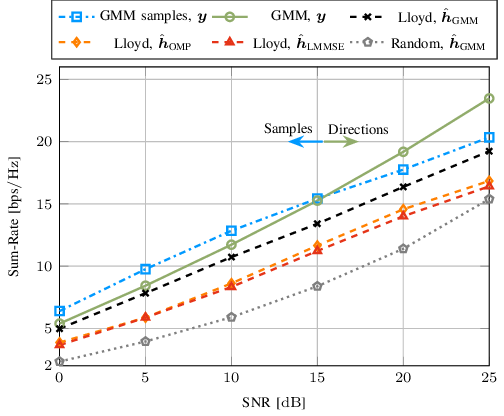}
    \caption{The average sum-rate ($\Ntx=16$, $\Nrx=4$, and $J=4$ \acp{MT}) over the \ac{SNR} achieved with different feedback approaches when the iterative \ac{WMMSE} or the \ac{SWMMSE} are employed, for $n_p=8$ pilots, and $B=6$ bits.}
    \label{fig:MUcbatSNR0PX_SNRXP8_WMMSEd1_new}
\end{figure}

In \Cref{fig:MUcbatSNR0PX_SNRXP8_WMMSEd1_new}, we still consider a setting with $n_p=8$ pilots but vary the \ac{SNR}.
We depict the sum-rate averaged over all constellations.
We can see, that our proposed \ac{GMM}-based feedback approach, with either exploiting directional information (``GMM, $\mby$'') or the generative modeling-based approach (``GMM samples, $\mby$'') outperform the conventional approaches.
We can observe, that in this case, up to an \ac{SNR} of $\approx\SI{15}{dB}$, the generative modeling-based method performs better, and for larger \ac{SNR} values, the directional approach is superior.
This is illustrated by the arrows in \Cref{fig:MUcbatSNR0PX_SNRXP8_WMMSEd1_new}.
Thus, the results suggest that jointly designing the precoders by solving the ergodic sum-rate maximization problem from \eqref{eq:ergodic_sumrate} by utilizing the \ac{SWMMSE}, is beneficial for low to medium \ac{SNR} values, and the directional approach, which exploits the iterative \ac{WMMSE}, is superior for larger \ac{SNR} values.

\begin{figure}[tb]
    \centering
    \includegraphics[scale=1.0]{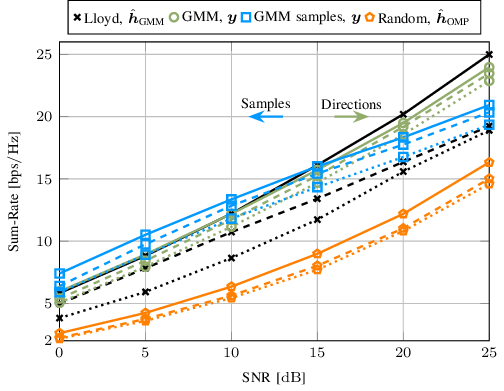}
    \caption{The average sum-rate ($\Ntx=16$, $\Nrx=4$, and $J=4$ \acp{MT}) over the \ac{SNR} achieved with different feedback approaches using the iterative \ac{WMMSE} or the \ac{SWMMSE}, for different number of pilots $n_p$ (dotted: $n_p=2$, dashed: $n_p=8$, solid: $n_p=16$), and $B=6$ bits.}
    \label{fig:MUcbatSNR0PX_SNRXPXX_WMMSEdBest}
\end{figure}

This observation is also supported by the results in \Cref{fig:MUcbatSNR0PX_SNRXPXX_WMMSEdBest}, where we depict the performances of our proposed approaches ``GMM, $\mby$'' and ``GMM samples, $\mby$'' and compare them to the best performing conventional method ``Lloyd, $\hhat_{\text{GMM}}$'', and the random codebook-based approach which uses the \ac{OMP} estimator ``Random, $\hhat_{\text{OMP}}$'' (i.e., no environment awareness) for a varying \ac{SNR} and $n_p \in \{2,8,16\}$.
There, dotted curves represent $n_p=2$, dashed curves $n_p=8$, and solid curves $n_p=16$ pilots.
The approaches with no environment awareness, i.e., ``Random, $\hhat_{\text{OMP}}$'' perform worst.
Both of our proposed approaches, i.e., ``GMM, $\mby$'' and ``GMM samples, $\mby$'', with only $n_p=2$ pilots, outperform the conventional ``Lloyd, $\hhat_{\text{GMM}}$'' approach with four times more, i.e., $n_p=8$, pilots.
When the number of pilots is equal to the number of transmit antennas (large pilot overhead), i.e., $n_p=\Ntx=16$, our proposed approaches which solely require the \ac{GMM} at the \acp{MT}, at least can compete with the conventional ``Lloyd, $\hhat_{\text{GMM}}$'' approach, which requires both the \ac{GMM} and the Lloyd directional codebook at each \ac{MT}.
For \acp{SNR} up to about $\SI{15}{dB}$ our proposed generative modeling-based approach even outperforms the conventional method based on the Lloyd directional codebook.

\begin{figure}[tb]
    \centering
    \includegraphics[scale=1.0]{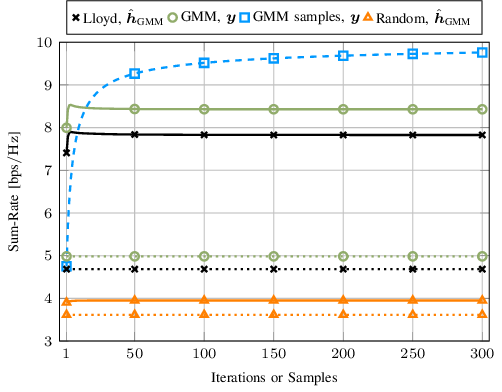}
    \caption{The average sum-rate ($\Ntx=16$, $\Nrx=4$, and $J=4$ \acp{MT}) over the number of iterations/samples achieved with different feedback approaches and precoding techniques (dotted: RBD (remains constant, since non-iterative), dashed: SWMMSE, solid: WMMSE) for a system with an $\text{SNR}=\SI{5}{dB}$, $n_p=8$ pilots, and $B=6$ bits.}
    \label{fig:MUcbatSNR5P8overiter_WMMSEd1CMPtoRBDandRCI}
\end{figure}

Finally, in \Cref{fig:MUcbatSNR5P8overiter_WMMSEd1CMPtoRBDandRCI}, we present how the sum-rate evolves over the number of iterations in the case of the iterative \ac{WMMSE} (solid curves), or over the number of drawn samples (per \ac{MT}) in the case of the \ac{SWMMSE} (dashed curves) for a setting with an $\text{SNR}=\SI{5}{dB}$ and $n_p=8$ pilots.
In comparison, we depict the performance of the case, where we applied \ac{RBD} (dotted curves) in order to jointly design the precoders.
Note that since \ac{RBD} is a non-iterative approach, the respective curves are constant over the iterations.
We can observe that in the case of random codebooks (``Random, $\hhat_{\text{GMM}}$'') the performance gains achieved by using the iterative \ac{WMMSE}, compared to using \ac{RBD}, are relatively small.
In contrast, using the Lloyd (``Lloyd, $\hhat_{\text{GMM}}$'') or \ac{GMM} (``GMM, $\mby$'') directional codebook approaches, we obtain huge performance gains when we use the iterative precoding techniques.
In these cases, already a small number of iterations is enough to reach the performance maximum.
Interestingly, a small overshoot can be observed.
This artefact is possibly due to the fact that the iterative \ac{WMMSE} is designed for perfect \ac{CSI}, but here we are restricted to using directional information due to the limited feedback.
In contrast, when we use the generative modeling-based approach (``GMM samples, $\mby$'') we can observe that the performance steadily improves over the number of drawn samples.
We observed this behavior consistently for different \ac{SNR} values and numbers of pilots.

\section{Conclusion}
\label{sec:conclusion}

In this work, we have investigated a novel \ac{GMM}-based feedback scheme for \ac{FDD} systems.
In particular, we proposed to use a \ac{GMM} for codebook construction, feedback encoding, and as a generator, which provides statistical information about the channels of the \acp{MT} to the \ac{BS}.
The proposed scheme exhibits lower computational complexity as compared to state-of-the-art approaches, and even allows for parallelization at the \acp{MT}.
Moreover, the proposed scheme stands out through its versatility.
That is, given the feedback information of the \acp{MT} at the \ac{BS}, it is flexible in deciding for the single-user or the multi-user transmission mode.
The versatility is even more pronounced through a convenient adaption at the \acp{MT} to any desired \ac{SNR} and pilot configuration without retraining the \ac{GMM}.
This is a huge advantage as compared to existing end-to-end \ac{DNN} approaches, which do not provide this versatility.
Numerical results have demonstrated that the proposed feedback scheme outperforms conventional methods, especially in configurations with reduced pilot overhead.
The achieved performance gains of the proposed scheme can be leveraged to deploy systems with lower pilot overhead or even fewer feedback bits as compared to state-of-the-art methods.

\balance
\bibliographystyle{IEEEtran}
\bibliography{IEEEabrv,biblio}
\end{document}